\documentclass[12pt]{article}
 \pdfoutput=1
\usepackage{amssymb}
\usepackage{amsmath,mathtools}
\usepackage{amstext}
\usepackage{graphicx,epsfig}
\usepackage{epsfig}
\usepackage{verbatim} 
\usepackage{fancyhdr}
\usepackage{fancybox}
\usepackage{color}
\usepackage{ulem,bbold}
\usepackage{enumitem}
\usepackage{subfigure}
\usepackage{bbm}
\usepackage{multirow}
\usepackage{hhline}
\usepackage[table]{xcolor}
\usepackage{parskip}

\allowdisplaybreaks[1]

\newcommand{\be}{\begin{equation}}
\newcommand{\ee}{\end{equation}}
\newcommand{\bea}{\begin{eqnarray}}
\newcommand{\eea}{\end{eqnarray}}
\newcommand{\beas}{\begin{eqnarray*}}
\newcommand{\eeas}{\end{eqnarray*}}
\newcommand{\nn}{\nonumber}

\def\({\left(}
\def\){\right)}

\newcommand{\half}{\frac{1}{2}}

\newcommand{\Comment}[1]{{}}
\definecolor{MyDarkBlue}{rgb}{0.15,0.15,0.45}
\usepackage[linktocpage=true]{hyperref}
\hypersetup{
colorlinks=true,
citecolor=MyDarkBlue,
linkcolor=MyDarkBlue,
urlcolor=MyDarkBlue,
}

\newlength{\arrayrulewidthOriginal}

\newlength{\fdagwidth}
\newlength{\diagupwidth}
\newlength{\stepback}

\numberwithin{equation}{section}

\begin{document}

\begin{center}
{\Large \bf{Kaluza-Klein Towers on General Manifolds}}
\end{center} 
 \vspace{1truecm}
\thispagestyle{empty} \centerline{
{\large  {Kurt Hinterbichler${}^{a,}$}}\footnote{E-mail address: \Comment{\href{mailto:khinterbichler@perimeterinstitute.ca}}{\tt khinterbichler@perimeterinstitute.ca}}, \large{Janna Levin ${}^{b,c,}$}\footnote{E-mail address: \Comment{\href{mailto:janna@astro.columbia.edu}}{\tt janna@astro.columbia.edu}}
{\large  {and Claire Zukowski${}^{d,e,f,}$}}\footnote{E-mail address: \Comment{\href{mailto:czukowski@berkeley.edu}}{\tt czukowski@berkeley.edu}}
                                                          }

\vspace{1cm}

\centerline{{\it ${}^a$ 
Perimeter Institute for Theoretical Physics,}}
 \centerline{{\it 31 Caroline St. N, Waterloo, Ontario, Canada, N2L 2Y5}} 
 
 \centerline{{\it ${}^b$ 
Institute for Strings, Cosmology and Astroparticle Physics,}}
\centerline{{\it Physics Department, Columbia University, New York, NY 10027}}
 
 \centerline{{\it ${}^c$ 
Department of Physics and Astronomy, Barnard College, New York, NY 10027}}

\centerline{{\it ${}^d$ 
Center for Theoretical Physics and Department of Physics,}}
\centerline{{ \it University of California, Berkeley, CA 94720, U.S.A}}

\centerline{{\it ${}^e$ 
Lawrence Berkeley National Laboratory, Berkeley, CA 94720, U.S.A}}

\centerline{{\it ${}^f$ 
Kavli Institute for the Physics and Mathematics of the Universe (WPI),}} 
\centerline{{\it The University of Tokyo, Kashiwa, Chiba 277-8583, Japan}}

\begin{abstract}

A higher-dimensional universe with compactified extra dimensions admits a four-dimensional description consisting of an infinite Kaluza-Klein tower of fields.  We revisit the problem of describing the free part of the complete Kaluza-Klein tower of gauge fields, $p$-forms, gravity, and flux compactifications.  In contrast to previous studies, we work with a generic internal manifold of any dimension, completely at the level of the action, in a gauge invariant formulation, and without resorting to the equations of motion or analysis of propagators.  We demonstrate that the physical fields and St\"uckelberg fields are naturally described by ingredients of the Hodge decomposition and its analog for symmetric tensors.  The spectrum of states and stability conditions, in terms of the eigenvalues of various Laplacians on the internal manifold, is easily read from the action.

\end{abstract}

\newpage

\thispagestyle{empty}
\tableofcontents
\thispagestyle{empty}
\newpage
\setcounter{page}{1}
\setcounter{footnote}{0}

\section{Introduction}
\parskip=5pt
\normalsize
\vspace{1cm}

The universe may conceal curled-up
extra dimensions, a topic of fascination since the early work of Kaluza
and Klein \cite{Kaluza:1921tu,Klein:1926tv}. Beyond the abstract
curiosity of a higher-dimensional universe, there are real
implications for our four-dimensional experience if the
hidden dimensions are correspondingly real.  
Extra dimensions, if they exist, might be too small to explore directly,
but their existence may be inferred from patterns
imprinted on four-dimensional physics. 
Integration of the fields' actions over the internal manifold reduces
the higher-dimensional laws of physics to effective
four-dimensional laws of physics.
In addition to the four-dimensional counterparts of the fields, 
infinite towers of particles 
of increasing mass appear. 
The masses are given by the eigenvalues of appropriate Laplacians acting on
the internal space. 

There have been many studies of the Kaluza-Klein spectrum associated
with various compactifications, see for instance some of the classic
papers \cite{Salam:1981xd}, and some of the famous early studies of
supergravity compactifications
\cite{vanNieuwenhuizen:1984iz,Kim:1985ez}.   Here we revisit
the derivation of the Kaluza-Klein tower in
terms of the eigenvalues of appropriate internal Laplacians. 
We are motivated to do so by the desire for a complete reference that can later be applied to many possible physical
applications, and
because we know of no prior treatment that
a) applies for any internal manifold (not just spheres or tori or
symmetric spaces) of any dimension, b) proceeds entirely at the level of the action,
without resort to component-by-component analyses of the equations
of motion or propagators, c) is completely gauge invariant and does not require
imposition of gauge conditions for infinite towers of gauge fields and
gravity, and d) takes account of all subtleties, such as those
associated with zero modes, Killing vectors and conformal Killing
vectors of the internal space.

The goal of this paper is to describe such a treatment.   We cover all the bosonic cases of most interest: scalars, $p$-forms, gravity and flux-compactifications.  We include even the simplest cases, in an attempt to provide a
complete, cohesive, self-contained reference.
Most
important, some physical details are more transparent in the general
methodology advocated here. 
In
particular, maintenance of the gauge symmetry is enlightening. With
gauge symmetry intact, 
the higher harmonics of the various
Laplacians naturally form St\"uckelberg fields
\cite{Stueckelberg:1957zz} for the towers of gauge
symmetries.\footnote{See \cite{Aulakh:1985un,Maheshwari:1985ux} for
  earlier studies of the original $4+1$ dimensional Kaluza-Klein
  theory along these lines, and e.g. \cite{Dolan:1984fm} for more complicated theories.}

Crucial is the combined Hodge and eigenspace decomposition of fields
on the internal manifold, which we discuss in detail in the
Appendices.  The decompositions provide the natural basis to 
cleave physical fields from St\"uckelberg fields.  
Derivation of the four-dimensional Kaluza-Klein action is then a straightforward
integration of the decomposed fields over the
extra dimensions.  The integration is instant since the decomposition
is naturally orthonormal. The resulting action is straightforwardly written
as a tower of gauge invariant combinations of fields and their
St\"uckelbergs.  The mass spectrum is readily read from the action.

We treat here the case of free fields: the scalar in Section~\ref{scalarsection}, the Maxwell field in Section~\ref{vectorsection}, the abelian $p$-form in Section~\ref{pformsection},  the free graviton in Section~\ref{gravitonsection}, and in Section~\ref{fluxsection} the case of flux compactifications, where there is both a $p$-form and a graviton with non-trivial mixing.  We do not put any constraints on the product manifolds, other than those needed for consistent propagation of the graviton.

In each case, once the spectrum of lower-dimensional fields is
attained we address the important question of stability. For a
compactification to be stable, the lower-dimensional spectrum must not contain ghosts (particles
with a wrong-sign kinetic term), or tachyons (particles with a
negative mass squared). In many cases,
stability is argued with purely geometrical theorems, such as
the Lichnerowicz bound, which apply to the spectrum of the Laplacian or
related operators. In other cases there may exist certain eigenvalues
that result in instabilities, narrowing down the parameter space of
stable compactifications.  In particular, we address the question of whether the Kaluza-Klein graviton masses on compactifications to de Sitter space can ever saturate or violate the Higuchi bound, and we find that they cannot.

\textbf{Conventions:}  We use mostly plus signature.  We are considering fields on a direct product of smooth manifolds $\mathcal{M}\times \mathcal{N}$, where ${\cal M} $ is a $d$-dimensional spacetime and ${\cal N} $ is a compact $N$-dimensional internal Riemannian manifold.  The total spacetime has dimension $D=d+N$.   The coordinates on the full product space are $X^A$, with $A,B,\cdots$ running over $D$ values, and the metric is $G_{AB}(X)$.   The coordinates on ${\cal M} $ are $x^\mu$, with $\mu,\nu,\cdots$ running over $d$ values, and the metric is $g_{\mu\nu}(x)$.   The coordinates on ${\cal N} $ are $y^m$ with $m,n,\cdots$ running over $N$ values, and the metric is $\gamma_{mn}(y)$.  ${\cal V}_{\cal N}=\int d^Ny\sqrt{\gamma}$ is the volume of ${\cal N}$.  The Riemann curvature is defined so that for a vector $V^\mu$, we have $[\nabla_\mu,\nabla_\nu]V^\rho=R^\rho_{\ \sigma\mu\nu}V^\sigma$.  The Ricci curvature is $R_{\mu\nu}=R^\rho_{\ \mu \rho\nu}$ and the Ricci scalar is $R=R^\mu_{\ \mu}$.   By an Einstein space, we mean any space with a metric satisfying $R_{\mu\nu}=k g_{\mu\nu}$ with $k$ constant.  Symmetrization, $(\cdots)$, and anti-symmetrization, $[\cdots]$, of $p$ indices are defined as the sum over (signed) permutations with a pre-factor of $1/p!$.

\section{Scalar\label{scalarsection}}

As a warm-up, and for completeness, we start with the simplest example of a higher dimensional theory: the free scalar. 
The higher dimensional action for a free massless scalar $\phi(X)$ on ${\cal M}\times {\cal N}$ is
\be S =-\frac{1}{2} \int d^{D}X \sqrt{-G}\partial_{A} \phi\, \partial^{A} \phi~.\label{scalaraction} \ee

We must expand the scalar in appropriate eigenfunctions of the internal manifold ${\cal N} $.  The appropriate expansion is the Hodge eigenvalue decomposition, reviewed in Appendix~\ref{appendixforms}.  In the case of scalars, it reads
\be \phi(x,y) = \sum_{a}\phi^{a}(x)\psi_{a}(y)+\frac{1}{\sqrt{{\cal V}_{\cal N}}}\phi^0(x)~, \label{scalareigen}\ee
where $\psi_{a}(y)$ are a basis orthonormal eigenvectors of the scalar Laplacian on ${\cal N} $, $\int d^Ny\sqrt{\gamma}\ \psi^{\ast a}\psi_b=\delta^{a}_{b}$, and $ \square_{(y)}\psi_a=-\lambda_a\psi_a$, with positive eigenvalues $\lambda_a>0$, labeled by $a$ including multiplicities.  The constant piece (where ${\cal V}_{\cal N}=\int d^Ny\sqrt{\gamma}$ is the volume of ${\cal N} $, ensuring proper normalization) takes account of the zero eigenvalue of the Laplacian.  

Although the basis $\psi_a$ can always be chosen to be real, we allow it to be complex, since this is often more convenient, e.g. the spherical harmonics on the sphere.  The original field $\phi$ is real, so we have the restriction $\phi^{a\ast}=(\eta_a^{-1})\phi^{\bar a}$ on the lower dimensional fields arising as coefficients in the expansion Eq.~\eqref{scalareigen}.  Here the bar on the index indicates some involution on the set of indices and $\eta_a$ a possible phase, e.g. $(\bar l,\bar m)=(l,-m)$, $\eta_{l,m}=(-1)^m$ for the standard spherical harmonics on the two-sphere (see Appendix~\ref{KKanssection} for more explanation).

Plugging Eq.~\eqref{scalareigen} into Eq.~\eqref{scalaraction} and integrating over $\mathcal{N} $, using the orthonormality of the positive eigenfunctions, and the fact that the constant (harmonic) mode is orthogonal to the $\psi^a$, we arrive at the $d$-dimensional action
\be S= \int d^{d}x \sqrt{-g}\left[-\frac{1}{2}\(\partial_{} \phi^0\)^2-\sum_a \frac{1}{2}\left(\left|\partial_{} \phi^a\right|^2+\lambda_a\left|\phi^a\right|^2 \right)\right]~.\ee
 
The spectrum is 
\begin{itemize}
\item One massless scalar,
\item One massive scalar for each eigenvector $\psi_a$ of the scalar Laplacian with positive eigenvalue $\lambda_a$, with mass $m^2=\lambda_a$.
\end{itemize} 
The spectrum is stable.  There are never tachyonic scalars with $m^2<0$ or ghosts with wrong-sign kinetic terms.

The simplest case is $N=1$, for which the internal manifold ${\cal N}_1$ is the circle.  Here the eigenfunctions can be chosen to be $\psi_a={1\over \sqrt{2\pi {\cal R}}}e^{iay/ {\cal R}}$, where ${\cal R}$ is the radius of the circle and $a$ ranges over all the integers, with $a=0$ the zero mode, and $\bar a=-a$.  This reproduces the Fourier decomposition of the field over the circle.  The eigenvalues in this case are $\lambda_a=a^2/ {\cal R}^2$, and the spectrum consists of the massless scalar zero mode and a tower of massive doublets with $m^2=a^2/ {\cal R}^2$.

\section{Vector\label{vectorsection}}

We now proceed to the simplest example of a theory with higher-dimensional gauge invariance: the abelian vector.
The higher-dimensional action for an abelian vector field $A_A(X)$ on ${\cal M}\times {\cal N}$ is the Maxwell action,
\be \label{maxwell} S= -\frac{1}{4} \int d^{D}X \sqrt{-G} F_{AB} F^{AB}~,\ee
where $F_{AB} \equiv \nabla_{A} A_{B} - \nabla_{B} A_{A}$ is the usual field strength.
The theory is invariant under the gauge transformations
\be \delta A_{A} =  \partial_{A} \Lambda\label{vectorgaugexform}~,\ee
where  $\Lambda(X)$ is an arbitrary spacetime function.  This higher-dimensional gauge symmetry will appear from the $d$-dimensional point of view as an infinite tower of gauge symmetries acting on the Kaluza-Klein tower.

We will use the Hodge decomposition theorem (see Appendix~\ref{appendixforms}) to decompose the field into parts which will have a clear physical interpretation from the lower dimensional point of view. 
We write the vector field as follows,
\begin{align}
A_{A}(x,y) = 
\left\{
\begin{array}{l} \vspace{.3cm}
\displaystyle \sum_{a} a_{\mu}^a(x) \psi_{a}(y)+\frac{1}{\sqrt{{\cal V}_{\cal N}}}c_{\mu,0}(x)~,\\ 
\displaystyle \sum_{a}a^{a}(x) \partial_n \psi_a(y)+\sum_i b^{i}(x) Y_{i,n}(y)+\sum_\alpha c^\alpha(x) Y_{\alpha,n}(y)~.
\end{array}\label{vectfdec}
\right.
\end{align}
Here $\psi_{a}$ are the eigenmodes of the scalar Laplacian on ${\cal N} $ with positive eigenvalues $\lambda_a$, the $Y_{i,n}$ are co-exact (which satisfy $\nabla^nY_{i,n}=0$) eigenvectors of the vector Laplacian on ${\cal N}$ with positive eigenvalues $\lambda_i$, and the $Y_{\alpha,n}$ are harmonic eigenvectors of the vector Laplacian on ${\cal N}$, i.e. those with eigenvalue zero under the vector Laplacian.  These all obey the conditions and orthogonality properties described in Appendix~\ref{appendixforms}. The field $A_A$ is real, so we have the restrictions $a_{\mu}^{a\ast}=(\eta_a^{-1})a_{\mu}^{\bar a}$, $a^{a\ast}=(\eta_a^{-1})a^{\bar a}$,\ \ $b^{i\ast}=(\eta_i^{-1})b^{\bar i}$, $ c^{\alpha\ast}= (\eta_\alpha^{-1})c^{\bar\alpha}$.

Plugging into Eq.~\eqref{maxwell} and integrating over $y$ using orthogonality of the various eigenspaces, we find
\bea\label{gaugeinvactionvec}
\centerline{$\displaystyle S = \int d^{d}x \sqrt{-g}\left[ - \frac{1}{4}f_{\mu \nu, 0}^2+\sum_{a}\left(-\frac{1}{4}\left| f_{\mu \nu}^a\right|^2 -\frac{\lambda_a}{2}   \left| a_{\mu}^a-\partial_\mu a^a\right|^2\right) \right.$}\nn\\
\centerline{$\displaystyle \left. -\frac{1}{2} \sum_\alpha \left|\partial_\mu c^\alpha\right|^2  -\frac{1}{2}\sum_i\left( \left|\partial_\mu b^i\right|^2 +\lambda_i\left|b^i\right|^2\right)\right]~,$} \label{vacprestuk1}
\eea
where $f_{\mu \nu, 0}\equiv \nabla_\mu c_{\nu,0}-\nabla_\nu c_{\mu,0}$ and $f_{\mu \nu}^a\equiv \nabla_\mu a_\nu^a-\nabla_\nu a_\mu^a$.

The gauge parameter can also be expanded over the scalar eigenfunctions,
\be \Lambda(x,y) = \sum_{a} \Lambda^a (x) \psi_a (y) + \frac{1}{\sqrt{{\cal V}_{\cal N}}}\Lambda_{0}(x)~. \ee
Decomposing Eq.~\eqref{vectorgaugexform} and equating coefficients, the original gauge symmetry reduces to an infinite tower of gauge symmetries,
\begin{align}
\delta  a_{\mu}^a &= \partial_\mu \Lambda^a~,  \ \ \ \ \ \ \delta b^i=0~,  \nn\\
\delta  c_{\mu,0} &= \partial_\mu \Lambda_0~,  \ \ \ \ \ \  \delta c^\alpha =0~, \nn\\
\delta  a^a &= \Lambda^a~.
\end{align}
The action Eq.~\eqref{vacprestuk1} is gauge invariant, as it must be because it is a rewriting of the original gauge invariant higher dimensional action and no gauge has been fixed.  We can express it in terms of the gauge invariant combination
\be \tilde a_\mu^a=a_{\mu}^a-\partial_\mu a^a \ee
as follows:
\beas
\centerline{$\displaystyle S = \int d^{d}x \sqrt{-g}\left[ - \frac{1}{4}f_{\mu \nu, 0}^2+\sum_{a}\left(-\frac{1}{4}\left|\tilde f_{\mu \nu}^a\right|^2 -\frac{\lambda_a}{2}   \left|\tilde a_{\mu}^a\right|^2\right) \right.$}\\
\centerline{$\displaystyle \left.-\frac{1}{2} \sum_\alpha \left|\partial_\mu c^\alpha\right|^2 -\frac{1}{2}\sum_i\left(\left| \partial_\mu b^i \right|^2+\lambda_i\left|b^i\right|^2\right)\right]~.$}
\eeas

The gauge symmetry $\Lambda^a$ is a St\"uckelberg symmetry\footnote{See Section~4 of \cite{Hinterbichler:2011tt} for a review of the St\"uckelberg formalism applied to massive vectors.} \cite{Stueckelberg:1957zz}.  We can fix it by setting the unitary gauge $a^a = 0$, which amounts to setting $\tilde a_\mu^a=a_\mu^a$, from which we recover the standard Proca action for massive vectors.  The scalars corresponding to the tower of exact one-forms on the internal manifold ${\cal N}$ have become the St\"uckelberg fields carrying the longitudinal component of the tower of massive vectors associated with the scalar harmonics on ${\cal N}$, and the harmonics of the higher dimensional gauge symmetry have become the St\"uckelberg symmetries.

The spectrum is now manifest, we have
\begin{itemize}
\item One massless vector,
\item One massive vector for each eigenvector of the scalar Laplacian with positive eigenvalue, with mass $m^2=\lambda_a$,
\item One massless scalar for each harmonic one-form,
\item One massive scalar for each co-exact eigenvector of the vector Laplacian, with mass $m^2=\lambda_i$.
\end{itemize} 
The spectrum is stable.  There are never tachyonic particles with $m^2<0$ or ghosts with wrong-sign kinetic terms.

In the simplest case $N=1$, where the internal manifold ${\cal N}_1$ is the circle, there are no co-exact one-forms so the $i$ index is empty and there are no massive scalars.  There is only one massless scalar, corresponding to the single harmonic one-form which is a constant vector over the circle, a massless vector corresponding to the harmonic scalar, and a tower of massive vector doublets, with masses $m^2=a^2/ {\cal R}^2$, $a=1,2,3,\cdots$, where ${\cal R}$ is the radius of the circle.

\section{$p$-form\label{pformsection}}

The vector field generalizes to a $p$-form.
The higher dimensional action for a $p$-form gauge field $A_{A_1\cdots A_{p}}(X)$ on ${\cal M}\times{\cal N}$ is
\be \label{pform} S= -\frac{1}{2(p+1)!} \int d^{D}X \sqrt{-G} F_{A_1\cdots A_{p+1}} F^{A_1\cdots A_{p+1}}~, \ee
where $F_{A_1\cdots A_{p+1}} =(p+1)\, \nabla_{[A_1} A_{A_2\cdots A_{p+1}]}$ is the field strength.
The theory is invariant under the gauge transformations
\be \delta A_{A_1\cdots A_p} = p\, \nabla_{[A_1} \Lambda_{A_2\cdots A_p]}~,\label{formgxform}\ee
where  $\Lambda_{A_1\cdots A_{p-1}}(X)$ is an arbitrary ($p-1$)-form. 

We use the Hodge decomposition theorem reviewed in Appendix \ref{appendixforms} to write the components of the form field in terms of exact, co-exact and harmonic forms,
\begin{align}
A_{A_1\cdots A_p}(x,y) = 
\left\{
\begin{array}{l} \vspace{0.2cm}
\displaystyle \sum_{i_0} a_{\mu_1\cdots \mu_p}^{i_0} Y_{i_0}+\frac{1}{\sqrt{{\cal V}_{\cal N}}}c_{\mu_1\cdots \mu_p}^{\alpha_0}~,\\
 \vspace{0.2cm}
\displaystyle \sum_{i_1} a_{\mu_1\cdots \mu_{p-1}}^{i_1}Y_{i_1,n}+\sum_{i_0} b_{\mu_1\cdots \mu_{p-1}}^{i_0}\(dY_{i_0}\)_{n}+\sum_{\alpha_1}c_{\mu_1\cdots \mu_{p-1}}^{\alpha_1} Y_{\alpha_1,n}~,\\
\displaystyle \sum_{i_2} a_{\mu_1\cdots \mu_{p-2}}^{i_2}Y_{i_2,n_1n_2}+\sum_{i_1} b_{\mu_1\cdots \mu_{p-2}}^{i_1}\(dY_{i_1}\)_{n_1n_2}+\sum_{\alpha_2}c_{\mu_1\cdots \mu_{p-2}}^{\alpha_2} Y_{\alpha_2,n_1n_2}~,\\
\displaystyle \ \ \ \ \vdots\\
\displaystyle \sum_{i_q} a_{\mu_1\cdots \mu_{p-q}}^{i_q}Y_{i_q,n_1\cdots n_q}+\sum_{i_{q-1}} b_{\mu_1\cdots \mu_{p-q}}^{i_{q-1}}\(dY_{i_{q-1}}\)_{n_1\cdots n_q}+\sum_{\alpha_q}c_{\mu_1\cdots \mu_{p-q}}^{\alpha_q} Y_{\alpha_q,n_1\cdots n_q}~,\\
\displaystyle \ \ \ \ \vdots\\
\displaystyle \sum_{i_p} a_{}^{i_p}Y_{i_p,n_1\cdots n_p}+\sum_{i_{p-1}} b_{}^{i_{p-1}}\(dY_{i_{p-1}}\)_{n_1\cdots n_p}+\sum_{\alpha_p}c_{}^{\alpha_p} Y_{\alpha_p,n_1\cdots n_p}~.\\
\end{array}
\right.\label{fluxdecomp}
\end{align}
Here, $i_q$ indexes a basis of  co-exact (transverse) $q$-forms $Y_{i_q,n_1\cdots n_q}$ which are eigenvalues of the Hodge Laplacian Eq.~\eqref{Hodgelapgen} with eigenvalue $\lambda_{i_q}$. 
The $\alpha_q$ index a basis of harmonic $q$-forms $Y_{\alpha_q,n_1\cdots n_q}$.  (Note that $i_0$ corresponds to the index $a$, and $\alpha_0$ corresponds to the single value $0$.  The $Y_{i_0}=\psi_a$ are just the positive scalar eigenvalues of the Laplacian.)  These satisfy the conditions and orthogonality properties described in Appendix~\ref{appendixforms}.  The field $A_{A_1\cdots A_p}$ is real, so we have the reality conditions $a_{\mu}^{i_q\ast}=(\eta_{i_q}^{-1})a_{\mu}^{\bar i_q}$,\ \ $b^{i_q\ast}=(\eta_{i_q}^{-1})b^{\bar i_q}$, $ c^{\alpha_q\ast}=(\eta_{\alpha_q}^{-1}) c^{\bar\alpha_q}$.

Plugging Eq.~\eqref{fluxdecomp} into Eq.~\eqref{pform} and integrating over $y$ using orthogonality of the various eigenspaces, we find the Lagrangian\footnote{It is easiest to start by writing
\be F_{A_1\cdots A_{p+1}}^2=F_{\mu_1\cdots \mu_{p+1}}^2+\cdots+\left(\begin{array}{c}p+1 \\q\end{array}\right) F_{\mu_1\cdots \mu_{p+1-q}n_1\cdots n_q}^2+\cdots +F_{n_1\cdots n_{p+1}}^2.\label{eq45int}\ee
Each term of Eq.~\eqref{eq45int} then becomes a line in \eqref{pformstuk}.}
\footnotesize
\begin{align} 
\frac{\mathcal{L}}{\sqrt{-g}}=-{1\over 2(p+1)!}&\left(\sum_{i_0}\left|f_{\mu_1\cdots\mu_{p+1}}^{i_0}\right|^2 +{f_{\mu_1\cdots\mu_{p+1}}^{\alpha_0}}^2 \right) \nn \\
-{1\over 2\, p!}&\left(\sum_{i_0}\lambda_{i_0}\left|a_{\mu_1\cdots\mu_p}^{i_0}+(-1)^p \(db^{i_0}\)_{\mu_1\cdots\mu_p}\right|^2+\sum_{i_1}\left|f_{\mu_1\cdots\mu_{p}}^{i_1}\right|^2 +\sum_{\alpha_1}\left|f_{\mu_1\cdots\mu_{p}}^{\alpha_1}\right|^2 \right) \nn\\
&\vdots \nn\\
-{1\over 2\, (p+1-q)!}\left(\sum_{i_{q-1}}\lambda_{i_{q-1}}\right.&\left.\left|a_{\mu_1\cdots\mu_{p+1-q}}^{i_{q-1}}+(-1)^{p+1-q} \(db^{i_{q-1}}\)_{\mu_1\cdots\mu_{p+1-q}}\right|^2 +\sum_{i_q}\left|f_{\mu_1\cdots\mu_{p+1-q}}^{i_q}\right|^2+\sum_{\alpha_q}\left|f_{\mu_1\cdots\mu_{p+1-q}}^{\alpha_q}\right|^2 \right)\nn\\
&\vdots \nn\\
-{1\over 2\, }&\left(\sum_{i_{p-1}}\lambda_{i_{p-1}}\left|a_{\mu_1}^{i_{p-1}}- \(db^{i_{p-1}}\)_{\mu_1}\right|^2+\sum_{i_p}\left|f_{\mu_1}^{i_p}\right|^2 +\sum_{\alpha_p}\left|f_{\mu_1}^{\alpha_p}\right|^2 \right) \nn\\
-{1\over 2}&\sum_{i_p}\lambda_{i_p}\left|a^{i_p}\right|^2~,\label{pformstuk}
\end{align}
\normalsize
where $f\equiv da$ is the field strength of the form with the corresponding index.

The gauge parameter can also be expanded over the eigenforms,
\begin{align}
\Lambda_{A_1\cdots A_{p-1}}(x,y) = 
\left\{
\begin{array}{l} \vspace{0.3cm}
\displaystyle \sum_{i_0} \Lambda_{\mu_1\cdots \mu_{p-1}}^{i_0} Y_{i_0}+\frac{1}{\sqrt{{\cal V}_{\cal N}}}\zeta_{\mu_1\cdots \mu_{p-1}}^{\alpha_0}~,\\
\displaystyle \sum_{i_1} \Lambda_{\mu_1\cdots \mu_{p-2}}^{i_1}Y_{i_1,n}+\sum_{i_0} \omega_{\mu_1\cdots \mu_{p-2}}^{i_0}\(dY_{i_0}\)_{n}+\sum_{\alpha_1}\zeta_{\mu_1\cdots \mu_{p-2}}^{\alpha_1} Y_{\alpha_1,n}~,\\
\displaystyle \ \ \ \ \vdots\\
\displaystyle \sum_{i_q} \Lambda_{\mu_1\cdots \mu_{p-1-q}}^{i_q}Y_{i_q,n_1\cdots n_q}+\sum_{i_{q-1}} \omega_{\mu_1\cdots \mu_{p-1-q}}^{i_{q-1}}\(dY_{i_{q-1}}\)_{n_1\cdots n_q}+\sum_{\alpha_q}\zeta_{\mu_1\cdots \mu_{p-1-q}}^{\alpha_q} Y_{\alpha_q,n_1\cdots n_q}~,\\
\displaystyle \ \ \ \ \vdots\\
\displaystyle \sum_{i_{p-1}} \Lambda_{}^{i_{p-1}}Y_{i_{p-1},n_1\cdots n_{p-1}}+\sum_{i_{p-2}} \omega_{}^{i_{p-2}}\(dY_{i_{p-2}}\)_{n_1\cdots n_{p-1}}+\sum_{\alpha_{p-1}}\zeta_{}^{\alpha_{p-1}} Y_{\alpha_{p-1},n_1\cdots n_{p-1}}~. \label{pformgaugeans}
\end{array}
\right.
\end{align}

Expanding the gauge transformation law Eq.~\eqref{formgxform} and equating coefficients, the component fields get the transformation laws

\begin{align}
\delta a^{i_0}_{\mu_1\cdots \mu_p}&=\(d\Lambda^{i_0}\)_{\mu_1\cdots\mu_p}~, && \vdots\nn\\
\delta c^{\alpha_0}_{\mu_1\cdots \mu_p}&=\(d\zeta^{\alpha_0}\)_{\mu_1\cdots\mu_p}~, & \delta a^{i_q}_{\mu_1\cdots \mu_{p-q}}&=\(d\Lambda^{i_q}\)_{\mu_1\cdots\mu_{p-q}}~,\nn \\ 
& & \delta b^{i_{q-1}}_{\mu_1\cdots \mu_{p-q}}&=\(d\omega^{i_{q-1}}\)_{\mu_1\cdots\mu_{p-q}}+(-1)^{p+q}\Lambda^{i_{q-1}}_{\mu_1\cdots \mu_{p-q}}~,\nn \\
&& \delta c^{\alpha_q}_{\mu_1\cdots \mu_{p-q}}&=\(d\zeta^{\alpha_q}\)_{\mu_1\cdots\mu_{p-q}}~,\nn \\
 \delta a^{i_1}_{\mu_1\cdots \mu_{p-1}}&=\(d\Lambda^{i_1}\)_{\mu_1\cdots\mu_{p-1}}~, && \vdots\nn\\
\delta b^{i_0}_{\mu_1\cdots\mu_{p-1}} &=\(d\omega^{i_0}\)_{\mu_1\cdots\mu_{p-1}}+(-1)^{p+1}\Lambda^{i_0}_{\mu_1\cdots \mu_{p-1}}~, & \delta a^{i_p}_{}&=0~,\nn\\
\delta c^{\alpha_1}_{\mu_1\cdots \mu_{p-1}}&=\(d\zeta^{\alpha_1}\)_{\mu_1\cdots\mu_{p-1}}~, & \delta b^{i_{p-1}}_{}&=\Lambda^{i_{p-1}}_{}~,\nn \\
& \vdots & \delta c^{\alpha_p}_{}&=0~.
\end{align}

We form the gauge invariant combinations
\bea
\tilde a_{\mu_1\cdots\mu_p}^{i_0}&=&a_{\mu_1\cdots\mu_p}^{i_0}+(-1)^p \(db^{i_0}\)_{\mu_1\cdots\mu_p}~,\nn\\
&\vdots&\nn\\
\tilde a_{\mu_1\cdots\mu_{p+1-q}}^{i_{q-1}}&=&a_{\mu_1\cdots\mu_{p+1-q}}^{i_{q-1}}+(-1)^{p+1-q} \(db^{i_{q-1}}\)_{\mu_1\cdots\mu_{p+1-q}}~, \nn\\
&\vdots& \nn\\
\tilde a_{\mu_1}^{i_{p-1}}&=&a_{\mu_1}^{i_{p-1}}- \(db^{i_{p-1}}\)_{\mu_1}~,
\eea
in terms of which the Lagrangian is

\footnotesize
\bea
\frac{\mathcal{L}}{\sqrt{-g}}&=&-{1\over 2(p+1)!}\left(\sum_{i_0}\left|\tilde f_{\mu_1\cdots\mu_{p+1}}^{i_0}\right|^2 +(p+1)\, \lambda_{i_0}\left|\tilde a_{\mu_1\cdots\mu_p}^{i_0}\right|^2+{f_{\mu_1\cdots\mu_{p+1}}^{\alpha_0}}^2 \right) \nn \\
&& -{1\over 2\, p!}\left(\sum_{i_1}\left|\tilde f_{\mu_1\cdots\mu_{p}}^{i_1}\right|^2 +p\,\lambda_{i_0}\left|\tilde a_{\mu_1\cdots\mu_p}^{i_0}\right|^2+\sum_{\alpha_1}\left|f_{\mu_1\cdots\mu_{p}}^{\alpha_1}\right|^2 \right) \nn\\
&\vdots &\nn\\
&& -{1\over 2\, (p+1-q)!}\left(\sum_{i_q}\left|\tilde f_{\mu_1\cdots\mu_{p+1-q}}^{i_q}\right|^2+(p+1-q)\lambda_{i_{q}}\right.\left.\left|\tilde a_{\mu_1\cdots\mu_{p-q}}^{i_{q}}\right|^2+\sum_{\alpha_q}\left|f_{\mu_1\cdots\mu_{p+1-q}}^{\alpha_q}\right|^2 \right)\nn\\
&\vdots& \nn\\
&&-{1\over 2\, }\left(\sum_{i_p}\left| f_{\mu_1}^{i_p}\right|^2+\lambda_{i_p}\left|a^{i_p}\right|^2 +\sum_{\alpha_p}\left|f_{\mu_1}^{\alpha_p}\right|^2 \right)~. \nn\\
\label{pformstuk2}
\eea
\normalsize

The gauge symmetries $\omega^{i_q}$ are higher order gauge symmetries, and can be absorbed by redefining $\tilde \Lambda^{i_q}=\Lambda^{i_q}+(-1)^{p+q+1}d\omega^{i_q}$.
Then the gauge symmetries $\tilde\Lambda^{i_q}$ are St\"uckelberg. We can fix them by setting $a^{i_q} = 0$, which amounts to setting $\tilde a^{i_q}=a^{i_q}$ ($q=0,\cdots,p-1$).

The spectrum is now manifest:
\begin{itemize}
\item Massless $q$-forms for each harmonic ($p-q$)-form, indexed by $\alpha_{p-q}$, $q=0,\ldots,p$.
\item Massive $q$-forms for each co-exact form ($p-q$)-form, indexed by $i_{p-q}$, $q=0,\ldots,p$ (including the positive eigenvalues of the scalar Laplacian at $q=p$), with mass $m^2=\lambda_{i_{p-q}}$.
\end{itemize} 
The spectrum is stable.  There are never tachyonic particles with $m^2<0$ or ghosts with wrong-sign kinetic terms.

Recall that in $d$ dimensions a massless $p$-form can be dualized into a ($d-p-2$)-form, and a massive $p$-form can be dualized into a ($d-p-1$)-form, so in any given example these dualities can be used to reformulate the $d$-dimensional action.
Note that many of the ingredients may be non-dynamical for low dimensions.  A massless $p$-form field in $d$ dimensions, and a ($d-p-2$)-form, which can be dualized to a $p$-form, are non-dynamical for $p\geq d-1$.  A massive $p$-form field in $d$ dimensions, and a ($d-p-1$)-form, which can be dualized to a $p$-form, are non-dynamical for $p\geq d$.

Hodge duality (reviewed in Appendix~\ref{hodgeappendix}) tells us that the spectrum of harmonic $p$-forms on ${\cal N}$ is identical to the spectrum of harmonic ($N-p$)-forms.  Thus the number of massless $q$-forms is the same the number of massless ($2p-N-q$)-forms.

In the simplest case $N=1$, where the internal manifold ${\cal N}_1$ is the circle, there are no co-exact one-forms so the $i$ indices are  empty and there are no massive forms of degree $<p$.  There is only one massless ($p-1$)-form, corresponding to the single harmonic one-form which is a constant vector over the circle, a massless $p$-form corresponding to the harmonic scalar, and a tower of massive $p$-form doublets, with masses $m^2=a^2/ {\cal R}^2$, $a=1,2,3,\cdots$, where ${\cal R}$ is the radius of the circle.

\section{Graviton\label{gravitonsection}}

In the cases of the scalars, vectors, and $p$-forms, we were free to choose the background Kaluza-Klein manifold as we pleased.  
This is no longer the case for a graviton.  For a graviton to consistently propagate on a background spacetime, that background must be a solution to Einstein's equations~\cite{Aragone:1979bm, Deser:2006sq}.  Thus we must first find backgrounds in the form of a $D=(d+N)$-dimensional product space $\mathcal{M} \times\mathcal{N} $ which satisfy Einstein's equations.

\subsection{Background\label{gravbackround}}

The action for gravity in $D$ dimensions with a $D$-dimensional cosmological constant $\Lambda_{(D)}$ and $D$-dimensional Planck mass  $M_P$ is 
\be\label{EHaction} S= \frac{M_{P}^{D-2}}{2}\int d^D X \sqrt{-{G}} \left(R_{(D)}-2\Lambda_{(D)}\right)~. \ee
The Einstein equations for the metric are
\be R_{AB}-{1\over 2}{R_{(D)}}{G}_{AB}+\Lambda_{(D)}{G}_{AB}=0~.\ee
Taking the trace and solving for the Ricci curvature, these
can be equivalently written as
\be R_{AB}={R_{(D)}\over D}{G}_{AB}~,\ \ \  R_{(D)}={2D\over D-2}\Lambda_{(D)}~.\label{einsteintr}\ee
Breaking Eq.~\eqref{einsteintr} into its ${\cal M}$ components and ${\cal N}$ components, we find that both factors must be Einstein spaces,
\bea  R_{\mu\nu}&=&{R_{(d)}\over d}g_{\mu\nu}~,\ \ \  R_{(d)}\ {\rm constant}~, \\
 R_{mn}&=&{R_{(N)}\over N}\gamma_{mn}~,\ \ \  R_{(N)}\ {\rm constant}~, 
\eea
where the curvatures on ${\cal M}$ and $\cal{N}$ are given by,
\be R_{(d)}={2d\over d+N-2}\Lambda_{(D)}~,\ \ \  R_{(N)}={2N\over d+N-2}\Lambda_{(D)}~.\ee
We have the useful relations
\be {R_{(d)}\over d}={R_{(N)}\over N}~,\ \ \ \Lambda_{(D)} = \frac{1}{2}\left(1-\frac{1}{d}\right)R_{(d)} + \frac{1}{2}\left(1-\frac{1}{N}\right)R_{(N)}~.\label{curvarelate}\ee

(In the case $D=2$ (i.e. $d=N=1$), we must have $\Lambda_{(D)}=0$, but in this case gravity is topological and the action for fluctuations is a total derivative, so there is nothing to Kaluza-Klein reduce.)

\subsection{Linear Action}

We now write the full metric as
\be G_{AB} + \frac{2}{M_P^{\frac{D}{2}-1}}H_{AB}~, \ee
where $G_{AB}$ satisfies the background equations of Section~\ref{gravbackround}, and $H_{AB}$ is the fluctuation.  We expand the action Eq.~\eqref{EHaction} to second order in $H_{AB}$.
The result is the standard action for linearized gravity on a curved background,
\bea \nn S&=&\int d^DX\ \sqrt{-G}\left[ -{1\over 2}\nabla_C H_{AB} \nabla^C H^{AB}+\nabla_C H_{AB} \nabla^B H^{AC}-\nabla_A H_{(D)}\nabla_B H^{AB}+\half \nabla_A H_{(D)}\nabla^A H_{(D)}\right. \\ &&\left. +{R_{(D)}\over D}\left( H^{AB}H_{AB}-\half H_{(D)}^2\right)\right]~. \label{curvedmassivelin}\eea
Here the metric, covariant derivatives and curvature $R_{(D)}$ are those of the background, and satisfy the background equations of motion.  Indices are always moved with the background metric.

The linear action Eq.~\eqref{curvedmassivelin} is invariant under gauge transformations which are the linearized diffeomorphisms of GR,
\be \delta H_{AB}=\nabla_A\Xi_B+\nabla_B\Xi_A~,\label{fullvectorgtf}\ee
with $\Xi^A(X)$ the vector gauge parameter. 

\subsection{Reduction of Fluctuations}

We proceed to split the components of the metric fluctuation $H_{AB}$ into their lower dimensional pieces.  The components $H_{\mu\nu}$ are scalars in the internal dimensions, and should, like the scalar field in Section~\ref{scalarsection}, be split into eigenmodes of the scalar Laplacian.  The components $H_{\mu n}$ are vectors in the internal dimensions, and should be split according to the vector Hodge decomposition, analogously to the internal components of the vector field in Section~\ref{vectorsection}.  The new ingredient that does not appear in the case of the $p$-forms is the components $H_{mn}$, which are symmetric tensors in the extra dimensions.  The best way to split these is to use the symmetric tensor version of the Hodge decomposition, reviewed in Appendix~\ref{appendixtensors}. 

This leads to the following ansatz 
\bea \nn H_{\mu\nu}&=& \sum_a h_{\mu\nu}^a\psi_a+\frac{1}{\sqrt{{\cal V}_{\cal N}}}h_{\mu\nu}^0 \\ \nn
H_{\mu n}&=&\sum_i A_\mu^iY_{n,i}+\sum_a A_\mu^a\nabla_n\psi_a\\ \nn
H_{mn}&=&\sum_{\cal I} \phi^{\cal I} h_{mn,{\cal I}}^{TT}+\sum_{i\not={\rm Killing}}\phi^i\left(\nabla_m Y_{n,i}+\nabla_n Y_{m,i}\right)\\
\nn&&+\sum_{a\not={\rm conformal}} \phi^a\left(\nabla_m\nabla_n\psi_a-{1\over N}\nabla^2\psi_a \gamma_{mn}\right) +\sum_a {1\over N} \bar\phi^a\psi_a \gamma_{mn}+{1\over N}\frac{1}{\sqrt{\mathcal{V}_N}}\phi^0 \gamma_{mn}~. \\ \label{ansatzH}
\eea

Several comments are in order, starting at the top: 
the $H_{\mu\nu}$ components are split just as in Section~\ref{scalarsection}, the $\psi^a$ are positive orthonormal eigenmodes of the scalar Laplacian, and there is a zero mode $h_{\mu\nu}^0$.  In the split of the $H_{\mu n}$ components, the $Y_{n,i}$ are orthonormal co-closed (i.e. transverse) eigenvectors of the vector Laplacian.  In other words, in contrast to Eq.~\eqref{vectfdec} in the case of the vector field, here we have combined the co-exact and harmonic forms together in the single index $i$.  This is because the harmonic forms will play no special role in the case of pure gravity (as they did for the vector, where they correspond to massless scalars in $d$ dimensions), so they need not be indexed and carried around separately.  Instead, it is the Killing vectors which will be important, and will correspond to massless vector modes in $d$ dimensions.

Moving to the split of the $H_{m n}$ components, the $h_{mn,{\cal I}}^{TT}$ are symmetric transverse traceless orthonormal eigenfunctions of the Lichnerowicz operator Eq.~\eqref{lichoperatorA}, which is the natural Laplacian on the space of symmetric tensors on an Einstein space.   
Special care has been extended to those co-exact vectors that are Killing vectors (which we denote $i={\rm Killing}$), and those scalars that are conformal scalars (which we denote $a={\rm conformal}$).  The Killing vectors are precisely the co-exact one-forms that have eigenvalue $\lambda_{i}={2R_{(N)}\over N}$, which is the lowest possible eigenvalue for such forms (see Appendix~\ref{appendixkilling}, which collects facts about Killing vectors on closed Einstein manifolds).  Conformal scalars are those scalars whose gradients are conformal Killing vectors which are not Killing.  Conformal scalars exist only on the sphere, and are the scalar eigenfunctions with the lowest non-zero eigenvalue, $\lambda_a={R_{(N)}\over N-1}$ (see Appendix~\ref{appendixconformal}, which collects facts about conformal scalars on closed Einstein manifolds).
The $\phi^i$ scalars are not present in the decomposition when the index $i$ takes a value corresponding to an eigenmode which is a Killing vector.   Thus we explicitly exclude the Killing vectors from the sums in the expression for $H_{mn}$.  
We will find that there is $d$-dimensional vector field for each co-exact one-form labeled by $i$, with mass $m_i^2=\lambda_{i}-{2R_{(d)}\over d}$.  The vector is massless only for those $i$ which are Killing vectors, for which $\lambda_{i}={2R_{(N)}\over N}={2R_{(d)}\over d}$.  The $\phi^i$ will be the longitudinal mode of the massive vectors, and the longitudinal mode does not exist for the case when $i$ is a Killing vector, corresponding to massless vector in $d$ dimensions.
Similarly, the $\phi^a$ scalars are not present when the index $a$ takes a value corresponding to an eigenmode that is a conformal scalar.  Thus we explicitly exclude the conformal scalars from the sums in the expression for $H_{mn}$.  We will see that there is a tower of massive scalars corresponding to the positive eigenmodes of scalar Laplacian, with masses $m_a^2=\lambda_a-\frac{2R_{(d)}}{d}$.  There will be no such scalar, however, for the case where $a$ corresponds to a conformal scalar. 

Since $H_{AB}$ is real, we have the reality conditions $h_{\mu\nu}^{a\ast}=(\eta_a^{-1})h_{\mu\nu}^{\bar a}$, $A_{\mu}^{a\ast}=(\eta_a^{-1})A_{\mu}^{\bar a}$, $\phi^{a\ast}=(\eta_a^{-1})\phi^{\bar a}$, $\bar\phi^{a\ast}=(\eta_a^{-1})\bar\phi^{\bar a}$, $A_{\mu}^{i\ast}=(\eta_i^{-1})A_{\mu}^{\bar i}$,  $\phi^{i\ast}=(\eta_i^{-1})\phi^{\bar i}$,  $\phi^{{\cal I}\ast}=(\eta_{\cal I}^{-1})\phi^{\bar {\cal I}}$.

For the gauge parameters, we expand as
\bea \Xi_\mu&=& \sum_a \xi_{\mu}^a\psi_a+\frac{1}{\sqrt{{\cal V}_{\cal N}}}\xi_\mu^0~, \\
 \Xi_n&=&\sum_i \xi^iY_{n,i}+\sum_a \xi^a\nabla_n\psi_a~.
 \eea

Expanding the gauge transformation Eq.~\eqref{fullvectorgtf} and equating components, we find the following gauge transformations for the lower-dimensional fields,

\begin{align}
 \delta h_{\mu\nu}^a&=\nabla_\mu \xi_\nu^a+\nabla_\nu \xi_\mu^a~, &  \delta \phi^{\cal I}&=0~, \nn\\
 \delta h_{\mu\nu}^0&=\nabla_\mu \xi_\nu^0+\nabla_\nu \xi_\mu^0~, &  \delta \phi^i&=\xi^i~,  \ \ \ i\not={\rm Killing} \nn \\
 \delta A_\mu^i&=\nabla_\mu \xi^i, &  \delta \phi^a&=2\xi^a ~, \ \ \ a\not={\rm conformal} \nn\\
 \delta A_\mu^a&=\xi_\mu^a+\nabla_\mu \xi^a~, &  \delta \bar\phi^a&=-{2}\lambda_a\xi^a~,  \nn\\
& & \delta \phi^0&=0~.
\end{align}

First, we see that the field $\phi^i$ (which only exists when $i$ is not a Killing vector) is pure St\"uckelberg, and we can define the following gauge invariant combination
\be \tilde A_\mu^i=A_\mu^i-\partial_\mu \phi^i~, \ \ \ i\not={\rm Killing}~.\ee
Next, we see that the fields $A_\mu^a$, and one combination of $\phi^a,\bar\phi^a$ are pure St\"uckelberg, and it will be possible to gauge them away.  However, when the index $a$ is that of a conformal scalar, the field $\phi^a$ does not exist.  We must treat this case separately.  
First consider the case where $a$ is not conformal.  We can define the following convenient gauge invariant combinations,
\bea &&F^a={\lambda_a }\phi^a+\bar\phi^a~,\ \ \ \ \ \ \ \ a\not={\rm conformal}~, \\
&& \tilde h_{\mu\nu}^a=h_{\mu\nu}^a-\(\nabla_\mu A_\nu^a+\nabla_\nu A_\mu^a\)+\nabla_\mu\nabla_\nu \phi^a~, \ \ \ \ a\not={\rm conformal}~.
\eea
In the case where $a$ is conformal, $\phi^a$ does not exist, so there is no gauge invariant scalar combination.   Thus we have only
\bea
&& \tilde h_{\mu\nu}^a=h_{\mu\nu}^a-\(\nabla_\mu A_\nu^a+\nabla_\nu A_\mu^a\)-{1\over \lambda_a}\nabla_\mu\nabla_\nu \bar\phi^a~, \ \ \ \ a={\rm conformal}~.
\eea
The gauge invariant scalar $F^a$ exists only for the non-conformal modes.

We are now ready to perform the decomposition.  We express our result in terms of $\epsilon(h)$, which denotes the $d$-dimensional massless graviton action,
\bea \epsilon(h)= -{1\over 2}\nabla_\lambda h_{\mu\nu}^\ast \nabla^\lambda h^{\mu\nu}+\nabla_\lambda h_{\mu\nu} ^\ast \nabla^\nu h^{\mu\lambda}-\nabla_\mu h^\ast\nabla_\nu h^{\mu\nu}+\half \nabla_\mu h^\ast\nabla^\mu h +{R_{(d)}\over d}\left( h^{\mu\nu\ast}h_{\mu\nu}-\half |h|^2\right)+c.c.\nn\\ \label{epsilondef}
\eea
This kinetic term is gauge invariant, so that $\varepsilon(h^a)=\varepsilon(\tilde h^a)$.

Inserting the decomposition Eq.~\eqref{ansatzH} into Eq.~\eqref{curvedmassivelin}, and doing the integrals over the extra dimensions ${\cal N} $ using the orthogonality of the various parts of the decomposition Eq.~\eqref{ansatzH}, we find the action
\be S = \int d^dx \left(\mathcal{L}_0 + \mathcal{L}_a + \mathcal{L}_i + \mathcal{L}_{\cal I}\right)~, \ee
where
\bea
\nn \frac{\mathcal{L}_0}{\sqrt{-g}} &=& \varepsilon(h^0) + h^{\mu\nu,0}(\nabla_\mu \nabla_\nu \phi^0 - \Box \phi^0 g_{\mu\nu}) - \frac{R_{(d)}}{d}h^0 \phi^0\\
& & + \frac{1}{2}\left(1-\frac{1}{N}\right)(\partial \phi^0)^2 -\frac{1}{2}\left(1-\frac{2}{N}\right) \frac{R_{(d)}}{d}(\phi^0)^2~,\\
\nn \frac{\mathcal{L}_a}{\sqrt{-g}} &=&\sum_{a\not={\rm conformal}} \varepsilon(\tilde h^a) - \frac{1}{2}\lambda_a\left(|\tilde h_{\mu\nu}^a|^2 -|\tilde h^a|^2\right)\\
\nn & & + \frac{1}{2}\left\{\tilde h^{\mu\nu, a *}(\nabla_\mu \nabla_\nu F^a - \Box F^a g_{\mu\nu}) + \left[\left(1-\frac{1}{N}\right)\lambda_a - \frac{R_{(d)}}{d}\right] \tilde h^{a*} F^a + c.c.\right\}\\ 
& & + \frac{1}{2}\left(1-\frac{1}{N}\right)|\partial F^a|^2 + \frac{1}{2}\left(1-\frac{2}{N}\right)\left[\left(1-\frac{1}{N}\right)\lambda_a - \frac{R_{(d)}}{d} \right]|F^a|^2\nn \\
&&+\sum_{a={\rm conformal}} \varepsilon(\tilde h^a) - \frac{1}{2}\lambda_a |\tilde h_{\mu\nu}^a|^2 + \frac{1}{2}\lambda_a |\tilde h^a|^2~,\\
\frac{\mathcal{L}_i}{\sqrt{-g}} &=& \sum_{i\not={\rm Killing}}-\frac{1}{2}|\tilde F_{\mu\nu}^i|^2 - \left(\lambda_i - \frac{2R_{(d)}}{d}\right) |\tilde A_\mu^i|^2+ \sum_{i={\rm Killing}}-\frac{1}{2}| F_{\mu\nu}^i|^2~, \\
\frac{\mathcal{L}_{\cal I}}{\sqrt{-g}} &=&\sum_{\cal I}-\frac{1}{2}|\partial \phi^{\cal I}|^2 - \frac{1}{2}\left(\lambda_{\cal I}  - \frac{2R_{(d)}}{d}\right) |\phi^{\cal I}|^2~,
\eea
where $\tilde F_{\mu\nu}^i$, $F_{\mu\nu}^i$ are the standard Maxwell field strengths of the corresponding vectors.

This $d$-dimensional action is manifestly gauge invariant, since the massive modes are written completely in terms of the gauge invariant variables.  The $A_\mu^a$ and a gauge non-invariant combination of $\phi^a,\bar\phi^a$ become the longitudinal St\"uckelberg fields \cite{Stueckelberg:1957zz} for the tower of massive gravitons\footnote{See \cite{ArkaniHamed:2002sp} and Section~4 of \cite{Hinterbichler:2011tt} for reviews of the St\"uckelberg formalism applied to massive gravitons.}, and the $\phi^i$ become St\"uckelberg fields for a tower of massive vectors.  The higher harmonics of the higher-dimensional diffeomorphism symmetry becomes the St\"uckelberg symmetry.  
We can now, if we like, impose a unitary gauge, by setting $\phi^a=A_\mu^a=\phi^i=0$, after which the action reads the same as above with $\tilde h_{\mu\nu}\rightarrow h_{\mu\nu}^a$, $\tilde A_\mu^i\rightarrow A_\mu^i$ and $F^a\rightarrow \bar\phi^a$.

Part of the spectrum can now be read off.  From $\mathcal{L}_{\cal I}$ we find a tower of scalars, one for each transverse traceless tensor mode of the Lichnerowicz operator, with mass $m_{\cal I}^2=\lambda_{\cal I}- \frac{2R_{(d)}}{d}$. These are massless precisely when $\lambda_{\cal I}=\frac{2R_{(d)}}{d}$.  As reviewed in Appendix~\ref{ModspaceA}, this is precisely the case where $h^{TT}_{mn}$ corresponds to a deformation which preserves the condition that the internal metric be Einstein, and which cannot be undone by a diffeomorphism or change in volume, i.e. it corresponds to a direction in the moduli space of Einstein structures.
From $\mathcal{L}_i$ we find a tower of massive vectors, one for each non-Killing co-closed one-form labeled by $i$, with mass $m_i^2=\lambda_{i}-{2R_{(d)}\over d}$, and we find precisely one massless vector for each Killing vector.
The remaining parts $\mathcal{L}_0$ and $\mathcal{L}_a$ are mixed and must be diagonalized.

\subsection{Diagonalization}

We now diagonalize the $\mathcal{L}_0$ and $\mathcal{L}_a$ parts of the action.  We first assume $d\geq 3$.  There are subtleties for the lower dimensional cases where gravity is non-dynamical, so we will treat them separately.  

We start with $\mathcal{L}_0$.  The zero mode graviton-scalar terms are diagonalized by making the linearized conformal transformation\footnote{Under a conformal transformation 
\be h_{\mu\nu} = h^{'}_{\mu\nu} + \pi g_{\mu\nu}~, \ee
for any scalar $\pi$, the graviton action $\epsilon(h)$ in \eqref{epsilondef}, including the curvature terms, transforms as
\bea 
\nn 2\epsilon(h) &=& 2\epsilon(h')+ \left[ (d-2)h^{'\mu\nu\ast}(\nabla_\mu\nabla_\nu \pi - \Box \pi g_{\mu\nu}) + \frac{(d-1)(d-2)}{2}|\partial \pi|^2 - \frac{d-2}{d}R_{(d)} \left({h'}^\ast \pi +\frac{d}{2}|\pi|^2\right)+c.c.\right]~.
\eea
}
\be h^0_{\mu\nu} = h^{'0}_{\mu\nu} -\frac{1}{d-2} \phi^0 g_{\mu\nu}~. \ee
This gives 
\bea
\nn \frac{\mathcal{L}_0}{\sqrt{-g}} &=& \varepsilon(h^0) + \frac{d+N-2}{2N(d-2)}\left(-(\partial\phi^0)^2 + \frac{2R_{(d)}}{d}(\phi^0)^2\right)~.
\eea
This is the action for a massless graviton and a scalar of mass $m_0^2=- \frac{2R_{(d)}}{d}$.  This scalar is the volume modulus of the internal manifold.  In the positively curved case, the mass term is tachyonic and there is an instability.

Now turn to $\mathcal{L}_a$.  We can diagonalize the graviton-scalar cross-terms by the following transformation 
\be h^a_{\mu\nu} = h^{'a}_{\mu\nu} - \frac{d\left(1-\frac{1}{N}\right)\lambda_a - R_{(d)}}{d(d-1)\lambda_a-(d-2)R_{(d)}}F^a g_{\mu\nu} + \frac{d\left(1+\frac{d-2}{N}\right)}{d(d-1)\lambda_a-(d-2)R_{(d)}}\partial_\mu\partial_\nu F^a~, \ \ \ a\not={\rm conformal}~. \ee
Note that by the Lichnerowicz bound $\lambda_a \geq \frac{R_{(N)}}{N-1}$ (see Appendix~\ref{appendixconformal}) combined with the curvature relation Eq.~\eqref{curvarelate}, the denominator is never zero so this is always a valid transformation. 

The result in terms of $ h^{'a}_{\mu\nu}$ is
\bea
\nn \frac{\mathcal{L}_a}{\sqrt{-g}} &=&\varepsilon(\tilde h^a) - {1\over 2}\lambda_a \left(\frac{1}{2}|\tilde h_{\mu\nu}^a|^2 - |\tilde h^a|^2\right) \\ 
& & + \frac{\left(1+ \frac{d-2}{N}\right)\left(d\left(1-\frac{1}{N}\right)\lambda_a - R_{(d)}\right)}{2\left(d(d-1)\lambda_a + (d-2)R_{(d)}\right)}\left(-|\partial F^a|^2 - \left(\lambda_a - \frac{2R_{d}}{d}\right)|F^a|^2\right)~,\ \ \ a\not={\rm conformal}~. \nn\\
\eea
There is a Fierz-Pauli massive graviton $\tilde h^a_{\mu\nu}$ \cite{Fierz:1939ix}\footnote{See Section~2 of \cite{Hinterbichler:2011tt} for a review of the Fierz-Pauli action.} for each positive eigenvector with $m^2=\lambda_a$, and a massive scalar $F^a$ for each non-conformal positive eigenvector with $m^2=\lambda_a - \frac{2R_{d}}{d}$.  
Using the Lichnerowicz bound Eq.~\eqref{lichbound}, it is possible to see that the kinetic terms for the scalars are always positive, so the scalars are never ghostly.\\
 
\noindent {\bf \underline{$d=2$ case}}:\\
In two dimensions, $d=2$, the kinetic part of the graviton action, $\sqrt{-g}\epsilon(h)$ with $\epsilon(h)$ as in Eq.~\eqref{epsilondef}, is a total derivative so we may drop it. 
The remaining gravitons have no derivatives and are therefore auxiliary fields and we should solve for them through their equations of motion.

Starting with $\mathcal{L}_0$, the zero mode sector after reduction now has the Lagrangian
\begin{align}
\nn \frac{\mathcal{L}_0}{\sqrt{-g}} &=  h^{\mu\nu,0}(\nabla_\mu \nabla_\nu \phi^0 - \Box \phi^0 g_{\mu\nu}) - \frac{R_{(2)}}{2}h^0 \phi^0\\
& + \frac{1}{2}\left(1-\frac{1}{N}\right)(\partial \phi^0)^2 -\left[\frac{1}{2}\left(1-\frac{2}{N}\right)\frac{R_{(2)}}{2}\right](\phi^0)^2~.\
\end{align}
This is the linearization of a version of dilaton gravity in two dimensions~\cite{Grumiller:2002nm}.

Turning to $\mathcal{L}_a$, the part of the reduced action corresponding to scalar eigenfunctions of the Laplacian is now
\begin{align}
\nn\frac{\mathcal{L}_a}{\sqrt{-g}} &=\sum_{a\not={\rm conformal}}- \frac{1}{2}\lambda_a\left( |\tilde h_{\mu\nu}^a|^2 -|\tilde h^a|^2\right)\\
\nn & + \frac{1}{2}\left\{\tilde h^{\mu\nu, a *}(\nabla_\mu \nabla_\nu F^a - \Box F^a g_{\mu\nu}) + \left[\left(1-\frac{1}{N}\right)\lambda_a - \frac{R_{(d)}}{d}\right] \tilde h^{a*} F^a +c.c.\right\}\\ 
\nn&+\frac{1}{2}\left(1-\frac{1}{N}\right)|\partial F^a|^2 + \frac{1}{2}\left(1-\frac{2}{N}\right)\left[\left(1-\frac{1}{N}\right)\lambda_a-\frac{R_{(d)}}{d}\right]|F^a|^2\\
&+\sum_{a={\rm conformal}}  - \frac{1}{2}\lambda_a \left( |\tilde h_{\mu\nu}^a|^2 -|\tilde h^a|^2\right)~.
\end{align}

For the non-conformal part, varying with respect to $h^{\mu\nu,a}$ gives an expression for the non-dynamical graviton,
\be h_{\mu\nu}^a = \frac{1}{\lambda_a}\nabla_\mu\nabla_\nu F^a -\frac{1}{\lambda_a}\left[\left(1-\frac{1}{N}\right)\lambda_a-\frac{R_{(2)}}{2}\right]F^a g_{\mu\nu}~,\ \ a\not={\rm conformal}~.\ee
Substituting this back into the action, we find
\begin{equation} 
\frac{\mathcal{L}}{\sqrt{-g}}=\sum_{a\not={\rm conformal}} -\frac{1}{2}\left[\left(1-\frac{1}{N}\right)-\frac{R_{(2)}}{2\lambda_a}\right]\left(|\partial F^a|^2 +\left(\lambda_a-R_{(2)}\right)|F^a|^2\right)~.
\end{equation}
There is a tower of scalars with masses $m_a^2=\lambda_a-\frac{2R_{(d)}}{d}$, the same expression as in the $d\geq3$ case.  
Again the Lichnerowicz bound Eq.~\eqref{lichbound} guarantees that the kinetic term never has the wrong sign, i.e. there are no ghosts.

For the conformal scalars, $a={\rm conformal}$, the equations of motion give $h_{\mu\nu}^a = 0$, and the conformal part of the Lagrangian reduces to zero.

In summary, for $d=2$ the spectrum contains the same ingredients as the $d\geq3$ case, with the exception of the zero mode gravitons and the scalar corresponding to the volume modulus, which take the form of dilaton gravity when $d=2$.

\noindent {\bf \underline{$d=1$ case}}:\\  
The $d=1$ case is somewhat trivial: here $R_{(d)}=0$, and all the graviton kinetic terms, Fierz-Pauli mass terms, and vector kinetic terms vanish identically.  All that remains of ${\cal L}_0$ is a kinetic term for the volume modulus $\phi^0$.  All that remains of the graviton in ${\cal L}_a$ is a cross term $\sim h F$, so $h^a$ is a multiplier which sets $F^a=0$.  Nothing dynamical remains of ${\cal L}_i$, since the vector kinetic terms vanish.  Thus the spectrum for $d=1$ is just the zero mode and the scalars of ${\cal L}_{\cal I}$.

\subsection{Spectrum}

Collecting all the ingredients, we now summarize the Kaluza-Klein spectrum of pure gravity:
\begin{itemize}
\item One massless graviton (linearized dilaton gravity for $d=2$),
\item A tower of Fierz-Pauli massive gravitons, one for each eigenvector of the scalar Laplacian with eigenvalue $\lambda_a>0$, with masses $m_a^2=\lambda_a$,
\item A massless vector for each Killing vector,
\item A tower of massive vectors, one for each non-Killing co-exact one-form labeled by $i$, with mass $m_i^2=\lambda_{i}-{2R_{(d)}\over d}$, $i\not={\rm Killing}$,
\item One scalar for the volume modulus, with a curvature-dependent mass $m_0^2 =- \frac{2R_{(d)}}{d}$ (part of dilaton gravity for $d=2$),
\item A tower of massive scalars, one for each non-conformal eigenvalue of the scalar Laplacian, with masses $m_a^2=\lambda_a-\frac{2R_{(d)}}{d}$, $a\not={\rm conformal}$ (not dynamical for $d=1$),
\item A tower of scalars for the transverse traceless tensor modes of the Lichnerowicz operator, with mass $m_{\cal I}^2=\lambda_{\cal I}- \frac{2R_{(d)}}{d}$. They are massless when $\lambda_{\cal I}=\frac{2R_{(d)}}{d}$, i.e. for each modulus of the Einstein structure.
\end{itemize} 

Many of the ingredients are in fact non-dynamical for low dimensions.  For example, massless gravitons are not dynamical for $d\leq 3$, the massive gravitons and massless vectors are non-dynamical in $d\leq 2$, and the vector fields are non-dynamical for $d=1$.

Finally, let us comment on the case $N=1$, the original case of Kaluza and Klein, where ${\cal N} $ is the circle.  The extra dimensional metric has only one component $\gamma_{yy}=1$.  There are no co-exact vectors other than the single Killing vector which is constant around the circle. There are no transverse traceless tensors, and no non-conformal scalars (every vector is a conformal Killing vector for $N=1$).  The scalar eigenfunctions are simply the Fourier modes, $\psi_a={1\over \sqrt{2\pi {\cal R}}}e^{iay/ {\cal R}}$, where ${\cal R}$ is the radius of the circle and $a$ ranges over all the integers, with $a=0$ the zero mode, and the eigenvalues are $\lambda_a=a^2/ {\cal R}^2$.   In this case, the ansatz Eq.~\eqref{ansatzH} simplifies to a simple Fourier expansion,
\bea \nn    {H_{\mu\nu}}&=& \frac{1}{\sqrt{2\pi \mathcal{R}}}\left[h_{\mu\nu}^0+\sum_{a=1}^\infty \left( h_{\mu\nu}^ae^{iay/ {\cal R}}+c.c.\right)\right]~,\\ \nn
  {H_{\mu y}}&=&\frac{1}{\sqrt{2\pi \mathcal{R}}}\left[A_\mu^0+\sum_{a=1}^\infty  \left({ia\over {\cal R}}  A_\mu^a e^{iay/ {\cal R}}+c.c.  \right)\right]~,\\ \nn
 {H_{yy}}&=&\frac{1}{\sqrt{2\pi \mathcal{R}}}\left[\phi^0+\sum_{a=1}^\infty\left(   \bar \phi^ae^{iay/ {\cal R}}+c.c.\right)\right]~. \\ \label{ansatzHn1}
\eea
In this case, we recover a massless graviton $h_{\mu\nu}^0$, a massless vector $A_\mu^0$, the massless dilaton $\phi^0$, and a tower of massive graviton doublets $h_{\mu\nu}^a$ with longitudinal modes $A_\mu^a$, $\bar \phi^a$ and masses $m^2=a^2/ {\cal R}^2$.

\subsection{Stability}

The Kaluza-Klein spectrum of fluctuations contains the information required to determine stability of the compactification.
Since none of the fields are ever ghost-like, the threat to stability comes from fields that may have masses which become tachyonic.

\textbf{\underline{Gravitons}:} For gravitons on a flat or negatively curved Einstein space, $R_{(N)}\leq0$, instability occurs when the Fierz-Pauli mass squared is negative, which never happens, because the mass squared is given by the non-zero scalar eigenvalues $\lambda_a$ which are always positive.

For positively curved Einstein spaces, $R_{(N)}>0$, gravitons are stable as long as their masses are above the Higuchi bound \cite{Higuchi:1986py},
\be m^2\geq{d-2\over d(d-1)}R_{(d)}~.\ee
The bound of Lichnerowicz (see Appendix~\ref{Lichboundap}) tells us that on the internal manifold, the smallest non-zero eigenvalue $\lambda_a$ of the scalar Laplacian is bounded from below,
\be \lambda_a \geq {R_{(N)}\over N-1}~.\label{Rlichbound}\ee

Since the mass squared of the graviton is just the eigenvalue, $m^2=\lambda_a$, to violate or saturate the Higuchi bound we would need
\be \lambda_a \leq {d-2\over d(d-1)}R_{(d)}={d-2\over N(d-1)}R_{(N)}~,\ee
where we have used Eq.~\eqref{curvarelate}.
This is consistent with Eq.~\eqref{Rlichbound} only if ${N\over N-1}\leq {d-2\over d-1}$, which is impossible since ${N\over N-1}>1$ and ${d-2\over d-1}<1$.  Thus the gravitons are always stable.

A massive graviton that saturates the Higuchi bound develops a scalar gauge symmetry and propagates one less degree of freedom than a massive graviton above the Higuchi bound.  Such a graviton is known as a partially massless graviton \cite{Deser:2001pe,Deser:2001us,Deser:2001wx,Deser:2006zx}.  As of this writing, there is no known consistent self-interacting theory that contains a stable background propagating a partially massless graviton, and attempts to find one have run into obstructions \cite{Deser:2013bs,Deser:2013uy,deRham:2013wv,Fasiello:2013woa}.  Here, we see that Kaluza-Klein reductions are no exception: a partially massless graviton can never arise in a pure Kaluza-Klein expansion.  The removal of the scalars $\phi^a$ in the decomposition Eq.~\eqref{ansatzH} when there are conformal Killing vectors does not result in the removal of the longitudinal modes of the corresponding massive gravitons, it instead results in the removal of the corresponding scalars $F^a$.

\textbf{\underline{Vectors}:}  For vectors, instability occurs when the mass squared is negative.  The mass squared is given by $\lambda_{i}-{2R_{(d)}\over d}$, which can never be negative by the arguments of Appendix \ref{appendixkilling}.  Thus the vectors are never unstable.

\textbf{\underline{Scalars}:}   The scalars are the only potential sources of instabilities.  
A scalar is stable for flat and dS space if its mass squared is non-negative.  For AdS spaces, the mass squared may be negative as long as it satisfies the Breitenlohner-Friedman bound \cite{Breitenlohner:1982bm,Mezincescu:1984ev},
\be m^2\geq {d-1\over 4d}R_{(d)}~.\label{BFboundref}\ee
Our only condition on the $d$-dimensional space was that it be Einstein, which is more general than the maximally symmetric Minkowski, dS and AdS spaces for which these bounds strictly apply.  However, since we know of no more general bounds, we will take these as our criteria for stability.
\begin{itemize}
 \item First consider the zero mode scalar $\phi^0$ corresponding to the volume modulus, which has a mass squared $m^2=-{2R_{(d)}\over d}$.  This is tachyonic for positively curved manifolds, so compactifications of pure gravity to positively curved spaces, e.g. de Sitter, are always unstable.  For flat manifolds, it is massless.  For negatively curved spaces, $\phi^0$ is safely massive.  In the case $d=2$ we have dilaton gravity, which in fact propagates no local degrees of freedom~\cite{Kim:1994nq}, so we may call this stable in all cases if we are only concerned with local degrees of freedom.

\item Next consider the tower of non-conformal scalars $F^a$, present when $N\geq 2$, with masses $m^2=\lambda_a-{2R_{(d)}\over d}$.  Since $\lambda_a>0$, for negative or flat curvature these fields all have $m^2>0$.  For positive curvature, there may in principle be tachyonic instabilities coming from these scalars if the internal manifold has low lying eigenvalues that are too small. 

\item Finally, consider the scalars $\phi^{\cal I}$ coming from the eigenstates of the Lichnerowicz operator, which have masses $m^2=\lambda_{\cal I}-{2R_{(d)}\over d}$, which may potentially become tachyonic.  We know of no universal bounds on the spectrum of the Lichnerowicz operator, so we cannot say in general when a Kaluza-Klein compactification is stable.

There are known cases where some of the $\phi^{\cal I}$ are tachyonic.  For example, compactifying on a space that is a product of two spheres, there can be a Lichnerowicz zero mode which corresponds to blowing up one of the spheres while shrinking the other in such a way that the total volume is fixed~\cite{Duff:1984sv,Berkooz:1998qp,DeWolfe:2001nz}.  This corresponds in the large $d$ dimensions to a tachyon with mass $m^2=-{2R_{(d)}\over d}$, so the instability scale is of order the $d$-dimensional Hubble scale.

\end{itemize}

\textbf{\underline{Maximally symmetric spaces}:}   
We can say more in the case where the internal manifold is a quotient of a maximally symmetric space.  
In this case, $R_{mnpq}={R_{(N)}\over N(N-1)}\(\gamma_{mp}\gamma_{nq}-\gamma_{mq}\gamma_{np}\)$, and the Lichnerowicz operator on transverse traceless tensors becomes $\Delta_L h_{mn}=-\nabla^2 h_{mn}+{2R_{(N)}\over N-1}h_{m n}.$  For $R_{(N)}=0$, the spectrum is non-negative, and the zero modes are the massless scalar moduli.  
For $R_{(N)}<0$, i.e. quotients of hyperbolic space, the spectrum is positive; there are no massless modes for which $\lambda_\alpha={2R_{(N)}\over N}$ (there are no moduli, a consequence of Mostow-Prasad rigidity \cite{mostow1973strong,springerlink:10.1007/BF01418789}).  For $R_{(N)}>0$, i.e. spheres, the spectrum of the ``rough" Laplacian, $-\nabla^2$, on symmetric rank $r$ tensors on the unit $N$-sphere is $l(l+N-1)-r$, where $l=0,1,2,\cdots$ is an integer \cite{Rubin:1983be}.  Using this for $r=2$, restoring the radius of the sphere and adding the curvature parts, we find the Lichnerowicz spectrum $\lambda_\alpha=\left[l(l+N-1)-2\right]{R_{(N)}\over N(N-1)}+{2R_{(N)}\over N-1}$.  The masses of the scalars are then $m^2=\lambda_\alpha-{2R_{(N)}\over N}=l(l+N-1){R_{(N)}\over N(N-1)}>0$, so we have no tachyons or moduli from the $\phi^{\cal I}$ sector in reductions on spheres.

Moving to the $F^a$ scalars, the only dangerous case was $R_{(N)}>0$.  In the case the maximally symmetric internal manifold is a sphere and the scalar eigenvalues are $\lambda_a=\frac{l(l+N-1)}{N(N-1)}R_{(N)}$.  The $l=1$ modes are the conformal modes, and for $l\geq 2$ the eigenvalues are $>{2R_{(N)}\over N}={2R_{(d)}\over d}$, so none of the $F^a$ states are tachyonic.

\textbf{\underline{Summary of stability results}:} 
\begin{itemize} \item Compactifications with positive curvature are all unstable in the volume modulus. 
They are stable in the non-conformal scalars $F^a$ if and only if the scalar Laplacian has no eigenvalues below ${2R_{(N)}\over N}$.
They are stable in the Lichnerowicz scalars if and only if the spectrum of the Lichnerowicz operator has no eigenvalues below ${2R_{(N)}\over N}$.
\item Compactifications with zero or negative curvature are always stable in the volume modulus and in the scalars $F^a$.  They are stable in the Lichnerowicz scalars (and hence completely stable) if and only if the spectrum of the Lichnerowicz operator has no eigenvalues below ${2R_{(N)}\over N}$.  
\item Compactifications on quotients of maximally symmetric spaces are always stable in the non-conformal scalars and in the Lichnerowicz scalars.  
\end{itemize}
Recall that for pure gravity, the $d$-dimensional curvature $R_{(d)}$, the $N$-dimensional curvature $R_{(N)}$, and the higher-dimensional cosmological constant $\Lambda$ all have the same sign or, in the flat case, all vanish, so the above results can be read in terms of the $d$-dimensional curvature or in terms of the $D$-dimensional cosmological constant.

\section{Flux Compactifications\label{fluxsection}}

We now proceed to consider the case of Freund-Rubin-type flux compactifications \cite{Freund:1980xh}.  These arise from gravity plus a possible cosmological constant in a product space $\mathcal{M} \times\mathcal{N} $, together with an $N$-form flux wrapping the internal space.\footnote{For the case of an internal manifold that is a product manifold, with a collection of flux fields that individually wrap each component, see~\cite{Brown:2013mwa}.}

\subsection{Background\label{fluxbacksec}}

The action is Einstein gravity in $D$ dimensions with a  $D$-dimensional cosmological constant $\Lambda_{(D)} $ and $D$-dimensional Planck mass $M_P$, minimally coupled to an ($N-1$)-form potential $A_{A_1\cdots A_{N-1}}(X)$,

\be S= \frac{M_{P}^{D-2}}{2}\int d^D x \sqrt{-{G}} \left(R_{(D)}-2\Lambda_{(D)}  - \frac{1}{N!} {F}_{A_1...A_N}^2\right)~, \ee
where the $N$-form field strength is $ {F}_{A_1...A_N} = (d{A})_{A_1...A_N} = N \nabla_{[A_1} {A}_{A_2... A_N]}~. $

The Einstein equations of motion for the metric are
\be R_{AB} - \frac{1}{2} R_{(D)} G_{AB} + \Lambda_{(D)} G_{AB} = \frac{1}{(N-1)!}  F_{A A_2...A_{N}}  F_B^{\ \ A_2...A_{N}} - \frac{1}{2N!} G_{AB}  F_{A_1...A_N}^2~, \label{fluxback}\ee
and the equations of motion for the ($N-1$)-form are
\be\nabla^{A_1}{F}_{A_1A_2...A_N}=0~.\label{formeqofmotion}\ee

We allow the $N$-form flux to take values only in the internal space, ${F}_{\mu A_1...A_{N-1}} = 0$, and its value on the internal space is proportional to the volume form on $\mathcal{N} $,
\be  F_{n_1...n_N} = Q \epsilon_{n_1...n_N}~. \label{formansa}\ee
This ansatz automatically solves the ($N-1$)-form equation of motion Eq.~\eqref{formeqofmotion}.  The constant of proportionality $Q$ has mass dimension one.

We now plug the ansatz Eq.~\eqref{formansa} into Eq.~\eqref{fluxback}, take the trace and use it to solve for the Ricci tensor:
\be R_{AB}={R_{(D)}\over D}{G}_{AB}~,\ \ \  R_{(D)}={2D\over D-2}\Lambda_{(D)}+{d-N\over D-2}Q^2~.\label{einsteintrf}\ee
Breaking Eq.~\eqref{einsteintrf} into its $d$- and $N$-dimensional components, with metrics $g_{\mu\nu}$ and $\gamma_{mn}$ respectively, we find that both factors of the product space must be Einstein,
\bea
R_{\mu\nu} &=& \frac{R_{(d)}}{d} g_{\mu\nu}~,\\
R_{mn} &=& \frac{R_{(N)}}{N} \gamma_{mn}~, \label{fluxeinsteinpieces}
\eea
with the following relations for the scalar curvatures on ${\cal M}$ and ${\cal N}$,
\be R_{(d)}={2d\over d+N-2}\Lambda_{(D)}-{d(N-1)\over d+N-2}Q^2~,\ \ \  R_{(N)}={2N\over d+N-2}\Lambda_{(D)}+{N(d-1)\over d+N-2}Q^2~.\ee
These can be solved to obtain the higher dimensional cosmological constant in terms of the lower dimensional curvatures,
\begin{align} 
\Lambda_{(D)} &= \frac{1}{2}\left(1-\frac{1}{d}\right)R_{(d)} + \frac{1}{2}\left(1-\frac{1}{N}\right)R_{(N)}~,\label{curvrela}
\end{align}
and the flux in terms of the lower dimensional curvatures,
\begin{align} 
Q^2 &= \frac{R_{(N)}}{N} - \frac{R_{(d)}}{d}~.\label{fluxreln}
\end{align}
Note that when the flux is non-vanishing, the total background product space with metric $G_{AB}$ is not required to be Einstein. 
The relation Eq.~\eqref{fluxreln} will be the most useful during the Kaluza-Klein reduction process.  

Quantum mechanically, there is often a Dirac quantization condition imposed on the flux strength $Q^2$.  A $p$-form couples to $(p-1)$-branes through the worldvolume action $S=M_P^{\frac{D}{2}-1}e\int A$,  where $e$ is a coupling constant of dimension $p-(D-2)/2$.  Demanding that the path integral integrand $e^{iS}$ be independent of which choice of gauge potential is used to represent the background flux can lead to quantization conditions.
For example, when the internal manifold ${\cal N}$ has the topology of a sphere, the condition is \cite{Henneaux:1986ht},
\be Q={2\pi n\over M_P^{\frac{D}{2}-1}e {\cal V}_{\cal N}}~,\ \ \ n\in {\rm integers}~.\ee
The quantity $Q$ is really a flux density.  We may define the flux as 
$q\equiv Q{\cal V}_{\cal N}~,$
which has dimension $1-N$ and is the volume-independent quantity that is quantized in units of the fundamental constants.  Since we are only working classically, we will not impose these quantization conditions, though they should be kept in mind because it is through these conditions that quantum mechanics can restrict the class of allowed solutions.

\subsection{Fluctuations}

Now we write the full metric and form field respectively as
\begin{align} 
& G_{AB} + \frac{2}{M_P^{\frac{D}{2}-1}}H_{AB}~,\label{hfluxcannorm}\\
& A_{A_1...A_{N-1}} +\frac{1}{M_P^{\frac{D}{2}-1}} \tilde A_{A_1...A_{N-1}}~,
\end{align}
where $G_{AB}$ and $A_{A_1...A_{N-1}}$ solve the background equations of motion from Section~\ref{fluxbacksec}, and $H_{AB}$, $\tilde A_{A_1...A_{N-1}}$ are the small fluctuations of the metric and form respectively.
The field strength is then $ F_{A_1...A_N} +\frac{1}{M_P^{\frac{D}{2}-1}}\tilde F_{A_1...A_N}~,$
where $F_{A_1...A_N}$ is the background field strength and the fluctuation is $\tilde F_{A_1...A_N} = (d\tilde A)_{A_1...A_N} = N \nabla_{[ A_1} \tilde A_{A_2... A_N]}~. $

Expanding up to quadratic order in the fluctuations, and dropping linear terms which vanish identically by the equations of motion, the linearized action is

\bea \nn \frac{\mathcal{L}}{\sqrt{- G}} &=& -{1\over 2} \nabla_C H_{AB}  \nabla^C H^{AB}+ \nabla_C H_{AB}  \nabla^B H^{AC}- \nabla_A H_{(D)} \nabla_B H^{AB}+\half  \nabla_A H_{(D)} \nabla^A H_{(D)} \\ 
\nn && - \left(\frac{1}{2}R_{(D)} - \Lambda_{(D)}\right) \left(H^{AB}H_{AB} - \frac{1}{2} H_{(D)}^2\right) + 2 R^{AB} \left(H_A^{\ \ C} H_{BC} - \frac{1}{2} H_{AB} H_{(D)} \right) \\
\nn &&  +\frac{1}{2N!}\left(H_{AB}^2-\frac{1}{2}H_{(D)}^2\right) F_{A_1...A_N}^2- \frac{1}{(N-2)!}H^{A_1B_1}H^{A_2B_2}  F_{A_1...A_N}  F_{B_1 B_2}^{\ \ \ \ A_3...A_N} \\ \nn
&& +\frac{2}{(N-1)!}\left(\frac{1}{2}H_{(D)} H^{A_1 B_1} - H^{A_1}_{\ \ C} H^{C B_1}\right)  F_{A_1...A_N} F_{B_1}^{\ \ A_2...A_N}\\
\nn&&- \frac{1}{2N!}\tilde F_{A_1...A_N}^2 \\
&& +\frac{2}{(N-1)!}H^{A_1B_1} \tilde F_{A_1A_2...A_N} F_{B_1}^{\ A_2...A_N} - \frac{1}{N!} H_{(D)} \tilde F_{A_1...A_N}  F^{A_1...A_N}~. 
\label{quadratic}\eea
The first two lines reduce to Eq.~\eqref{curvedmassivelin}, the linearized action for the graviton with no flux, if it is additionally imposed that the full space be Einstein. 

\subsection{$d+N$ Split and Ans\"atze}

There are four different parts to the expression Eq.~\eqref{quadratic}: the kinetic terms for the graviton (the first line), the non-derivative quadratic terms for the graviton (the second, third and fourth lines), the kinetic terms for the form fluctuations (the fifth line),  and the mixing terms between graviton and flux (the sixth line).  

Using the value for the background flux and the definition of the Hodge star (reviewed in Appendix~\ref{hodgeappendix}), the mixing terms (the sixth line of  Eq.~\eqref{quadratic}) simplify to
\begin{align}
& (-1)^{N-1}Q\left[2 H^{\mu n}\nabla_\mu (\ast \tilde A)_n  + (-1)^N 2H^{\mu m}\nabla^n (\ast \tilde A_\mu)_{mn} + \left(H_{(N)} - H_{(d)}\right)\nabla^n (\ast \tilde A)_n\right], \label{mixingb}
\end{align}
where $H_{(N)}\equiv \gamma^{mn}H_{mn}$ and $H_{(d)}\equiv g^{\mu\nu}H_{\mu\nu}$,
and the  non-kinetic terms for the graviton (the second, third and fourth lines of Eq.~\eqref{quadratic}) reduce to
\begin{align}
\nn \frac{R_{(d)}}{d}\left( H_{\mu\nu}^2 - \frac{1}{2} H_{(d)}^2\right)+ \frac{2R_{(d)}}{d}\left(H_{\mu n}^2 - \frac{1}{2} H_{(d)} H_{(N)} \right) +\left(\frac{R_{(d)}}{d} + Q^2\right)H_{mn}^2 - \frac{1}{2} \left(\frac{R_{(d)}}{d} + 2Q^2\right)H_{(N)}^2~. \\
\end{align}

The ansatz for the graviton we will use is
\bea \nn H_{\mu\nu}&=& \sum_a h_{\mu\nu}^a\psi_a+\frac{1}{\sqrt{{\cal V}_{\cal N}}}h_{\mu\nu}^0~, \\ \nn
H_{\mu n}&=&\sum_i A_\mu^iY_{n,i}+\sum_a A_\mu^a\nabla_n\psi_a+\sum_\alpha A_\mu^\alpha Y_{n,\alpha}~, \\ \nn
H_{mn}&=&\sum_{\cal I} \phi^{\cal I} h_{mn,{\cal I}}^{TT}+\sum_{i\not={\rm Killing}}\phi^i\left(\nabla_m Y_{n,i}+\nabla_n Y_{m,i}\right)+\sum_{\alpha\not={\rm Killing}}\phi^\alpha\left(\nabla_m Y_{n,\alpha}+\nabla_n Y_{m,\alpha}\right)\\
\nn&&+\sum_{a\not={\rm conf.}} \phi^a\left(\nabla_m\nabla_n\psi_a-{1\over N}\nabla^2\psi_a \gamma_{mn}\right) +\sum_a {1\over N} \bar\phi^a\psi_a \gamma_{mn}+{1\over N}\frac{1}{\sqrt{\mathcal{V}_N}}\phi^0 \gamma_{mn}~. \\ \label{ansatzH2}
\eea
Here there is a difference compared to the pure graviton case Eq.~\eqref{ansatzH}.  We have separated the harmonic one-forms from the co-closed one-forms, so that $i$ now ranges only over the co-exact forms and $\alpha$ ranges over the harmonic forms (i.e. we are using the original Hodge decomposition for one-forms).  We do this because the fields corresponding to the harmonic forms will mix with those coming from the $p$-forms, and we will have to treat them separately.

There are three distinct cases we are attempting to treat simultaneously: a positively curved internal manifold, $R_{(N)}>0$, a negatively curved internal manifold $R_{(N)}<0$, and a flat internal manifold $R_{(N)}=0$.  In the positively curved case, there are no harmonic one-forms (see the arguments in Appendix~\ref{appendixkilling}), so the set of $\alpha$ is empty.  In the negatively curved case, there are no Killing vectors (see the arguments in Appendix~\ref{appendixkilling}), so the sets of $i\not={\rm Killing}$, $\alpha\not={\rm Killing}$ are the same as the sets of $i$, $\alpha$.  In the flat case, the Killing vectors and harmonic forms are one and the same, so the set of $\alpha\not={\rm Killing}$ is empty.

We expand the ($N-1$)-form fluctuation $\tilde A_{A_1...A_{N-1}}$ as in Eq.~\eqref{fluxdecomp}.  For reducing Eq.~\eqref{mixingb} we need to take the Hodge star of the relevant parts of the flux decomposition Eq.~\eqref{fluxdecomp},
\begin{align}
(\ast \tilde A)_n &= \sum_{i_{N-1}} a^{i_{N-1}} (\ast Y_{i_{N-1}})_n + \sum_{i_{N-2}} b^{i_{N-2}} (\ast d Y_{i_{N-2}})_n+ \sum_{\alpha_{N-1}} c^{\alpha_{N-1}} (\ast Y_{\alpha_{N-1}})_n~,\label{fluxgravdecomp1}\\
(\ast \tilde A_\mu)_{mn} &= \sum_{i_{N-2}} a_\mu^{i_{N-2}} (\ast Y_{i_{N-2}})_{mn} + \sum_{i_{N-3}} b_\mu^{i_{N-3}} (\ast d Y_{i_{N-3}})_{mn}+ \sum_{\alpha_{N-2}} c_\mu^{\alpha_{N-2}} (\ast Y_{\alpha_{N-2}})_{mn}~.\label{fluxgravdecomp2}
\end{align}

We will use some facts, reviewed in Appendix~\ref{hodgeappendix}, to map these basis vectors onto ones that have simple overlap with the eigenfunctions in the graviton decomposition.  The Hodge star provides an isomorphism between the space of harmonic $p$-forms and the space of harmonic ($N-p$)-forms.  Thus the indices $\alpha$ (which is $\alpha_1$) and $\alpha_{N-1}$ take values over the same set.  Similarly, the Hodge star provides an isomorphism between the space of co-exact $p$-forms and the space of exact ($N-p$)-forms, so the indices $a$ (which is $i_0$, the case $p=0$) and $i_{N-1}$ take values over the same set and the indices $i$ (which is $i_1$, the case $p=1$) and $i_{N-2}$ take values over the same set.  For $N>2$, 
we may define our bases as in Eqs.~\eqref{harmhodbasdu} and~\eqref{excohodrel} in Appendix~\ref{appendixforms}:
\begin{align} 
Y_{i_{N-1}} &= \frac{1}{\sqrt{\lambda_{a}}}\ast dY_{a}~,\label{hodgelowrelations1}\\
  Y_{i_{N-2}} &= \frac{1}{\sqrt{\lambda_{i}}}\ast dY_{i}~, \label{hodgelowrelations2}\\
  Y_{\alpha_{N-1}} &= \ast Y_{\alpha}~.\label{hodgelowrelations3}
\end{align}

For $N=2$, the Hodge star maps the space of harmonic one-forms labeled by $\alpha$ into itself. In this case, we cannot choose all the basis vectors in accord with Eq.~\eqref{hodgelowrelations3} (although Eqs.~\eqref{hodgelowrelations1} and~\eqref{hodgelowrelations1} need not be modified).  Instead, the index $\alpha$ of the $Y_{\alpha}$ is chosen so that it can be divided into imaginary self dual and imaginary anti-self dual sets, $\alpha_+$ and $\alpha_-$.  The sets $Y_{\alpha_\pm}$ then transform under the Hodge star as (see Appendix~\ref{appendixforms}),
\be\ast Y_{\alpha_\pm} = \pm i Y_{\alpha_\pm}~.\label{imselfd}\ee

The extreme case $N=1$, the case where the internal manifold is a circle and the $p$-form is just a scalar, goes through as well.  In this case, Eq.~\eqref{hodgelowrelations2} is empty, and Eq.~\eqref{hodgelowrelations3} is a simple identity between the single constant function and the single constant vector.  Eq.~\eqref{hodgelowrelations1} relates the space of scalar eigenfunctions to itself and so may have to be supplemented by an additional unitary transformation depending on the choice of basis, which may then be reabsorbed by a field redefinition of $a^a$.  We say more about the $N=1$ case at the end of Section \ref{specfluxsec}, and point out as we go along how the lower dimensional cases differ.

Renaming the labels on the coefficients in Eqs.~\eqref{fluxgravdecomp1} and~\eqref{fluxgravdecomp2}, 
\begin{align}
a^{i_{N-1}}\leftrightarrow a^a~,\ \ b^{i_{N-2}} \leftrightarrow b^i~, \ \  c^{\alpha_{N-1}}\leftrightarrow c^{\alpha}~, \ \ a_\mu^{i_{N-2}}&\leftrightarrow a_\mu^i~, 
\end{align}
and using the inverse of Eqs.~\eqref{hodgelowrelations1}-\eqref{hodgelowrelations3} obtained by taking the Hodge star of both sides, we can write Eq.~\eqref{fluxgravdecomp1} and Eq.~\eqref{fluxgravdecomp2} as
\begin{align}
&(\ast \tilde A)_n = (-1)^{N-1}\left[\sum_a \frac{1}{\sqrt{\lambda_a}}a^a (d\psi_a)_n + \sum_i \sqrt{\lambda_i}b^i Y_{i,n} + \sum_{\alpha}c^{\alpha}Y_{\alpha,n}\right]~, \label{tildesaeq}\\
&(\ast \tilde A_\mu)_{mn} = \sum_i \frac{1}{\sqrt{\lambda_i}} a_\mu^i (dY_i)_{mn} +...~, 
\end{align}
where the ellipsis denotes contributions that are orthogonal to all terms in the graviton decomposition, and hence do not contribute in Eq.~\eqref{mixingb}.

\subsection{Gauge Symmetries and Gauge Invariant Combinations\label{fluxgaugecombsec}}

As in Section~\ref{pformsection}, the theory is invariant under the ($N-2$)-form gauge transformations acting on the form field fluctuations,
\be \delta \tilde A_{A_1\cdots A_{N-1}} = (N-1)\, \nabla_{[A_1} \Lambda_{ A_2\cdots A_{N-1}]}~,\label{nformgauget}\ee
where  $\Lambda_{A_1\cdots A_{N-2}}(X)$ is an arbitrary ($N-2$)-form. 
As in Section~\ref{gravitonsection}, the diffeomorphism symmetries acting on the graviton fluctuations are
\be \delta H_{AB}=\nabla_A\Xi_B+\nabla_B\Xi_A~.\label{fullvectorgtf2}\ee
The diffeomorphism symmetries act on the form field fluctuations as well.  Acting on the field strength (which is gauge invariant under Eq.~\eqref{nformgauget} so that the transformation is unambiguous), we have 
\be \delta \tilde F_{A_1...A_N} = \mathcal{L}_\Xi  F_{A_1...A_N} \equiv \Xi^A  \nabla_A  F_{A_1...A_N} + ( \nabla_{A_1} \Xi^A)  F_{AA_2...A_N}+...+ ( \nabla_{A_N} \Xi^A)  F_{A_1...A_{N-1}A}~.\label{diffonforms} \ee

For the diffeomorphism symmetries, we expand the gauge parameter over the eigenforms,
\begin{align}
\Xi_\mu &= \sum_a \xi^{a}_{\mu} \psi_a + \frac{1}{\sqrt{{\cal V}_{\cal N}}}\xi^{0}_{\mu}~,\\
\Xi_n &= \sum_i \xi^i Y_{n,i} + \sum_a \xi^a \nabla_n \psi_a + \sum_\alpha \xi^\alpha Y_{n,\alpha}~.
\end{align}
By expanding out Eq.~\eqref{fullvectorgtf2} and equating coefficients, we find how the diffeomorphisms act on the lower dimensional fields in the graviton decomposition,
\begin{align}
 \delta h_{\mu\nu}^a&=\nabla_\mu \xi_\nu^a+\nabla_\nu \xi_\mu^a~, &  \delta \phi^{\cal I}&=0~, \nn\\
 \delta h_{\mu\nu}^0&=\nabla_\mu \xi_\nu^0+\nabla_\nu \xi_\mu^0~, &  \delta \phi^i&=\xi^i~, \ \ \ i\not={\rm Killing}~, \nn \\
 \delta A_\mu^i&=\partial_\mu \xi^i~, &  \delta \phi^a&=2\xi^a ~, \ \ \ a\not={\rm conformal}~, \nn\\
 \delta A_\mu^\alpha &=\partial_\mu \xi^\alpha~, &  \delta \bar\phi^a&=-{2}\lambda_a\xi^a~,  \nn\\
 \delta A_\mu^a&=\xi_\mu^a+\partial_\mu \xi^a~,  & \delta \phi^0&=0~, \nn\\ 
 & &  \delta \phi^\alpha&=\xi^\alpha~, \ \ \ \alpha\not={\rm Killing}~.  
\end{align}

We also have the ($N-1$)-form gauge transformations descending from Eq.~\eqref{nformgauget} with the expansion Eq.~\eqref{pformgaugeans},
\begin{align}
\delta a^{a} &=0~,\nn\\
\delta a_\mu^i&=\partial_\mu \Lambda^i~,\nn\\
\delta b^{i}&=\Lambda^i~,\nn\\
\delta c^{\alpha}&=0~.
\end{align}

Descending from Eq.~\eqref{diffonforms}, we find the following diffeomorphism transformations for the form fields\footnote{The factors of 2 are from the factor of 2 in the canonical normalization of $H_{AB}$ in Eq.~\eqref{hfluxcannorm}.},
\begin{align}
\delta a^a &= 2Q\sqrt{\lambda_a} \xi^a~,\nn\\
\delta(a_\mu^i-\partial_\mu b^i)  &=-\frac{2Q}{\sqrt{\lambda_i}}\partial_\mu \xi^i~,\nn\\
\delta c^\alpha &= 2Q \xi^\alpha~. \label{diffonfluxlowd}
\end{align}

We define the quantities,
\bea \tilde h_{\mu\nu}^a&=&\begin{cases} h_{\mu\nu}^a-\(\nabla_\mu A_\nu^a+\nabla_\nu A_\mu^a\)+\nabla_\mu\nabla_\nu \phi^a~, & a\not={\rm conformal}~, \label{gaugecombosstart}\\  h_{\mu\nu}^a-\(\nabla_\mu A_\nu^a+\nabla_\nu A_\mu^a\)-{1\over \lambda_a}\nabla_\mu\nabla_\nu \bar\phi^a~, & a={\rm conformal}~, \end{cases}\\
F^a&=&{\lambda_a }\phi^a+\bar\phi^a~, \ \ \ \ a\not={\rm conformal}~,\\
K^a &=&  \begin{cases} -Q\phi^a + \frac{1}{\sqrt{\lambda_a}}a^a,~&  a\not={\rm conformal}~, \\ a^a+{Q\over\sqrt{\lambda_a}}\bar\phi^a, & a ={\rm conformal}, \end{cases}~\\
\tilde A_\mu^i&=&\begin{cases} A_\mu^i-\partial_\mu \phi^i~, &  i\not={\rm Killing}~, \\ A_\mu^i, & i={\rm Killing}~, \end{cases}   \\
\tilde a_\mu^i &=& \begin{cases} a_\mu^i - \partial_\mu b^i + \frac{2Q}{\sqrt{\lambda_i}}\partial_\mu \phi^i~, & i\not={\rm Killing}~, \\  a_\mu^i - \partial_\mu b^i~,& i={\rm Killing}~,\end{cases}  \\
\tilde A_\mu^\alpha&=&\begin{cases} A_\mu^\alpha-\partial_\mu \phi^\alpha~, & \alpha\not={\rm Killing}~, \\ A_\mu^\alpha-{1\over 2Q}\partial_\mu c^\alpha~, & \alpha={\rm Killing}~, \end{cases}\label{alphaginv1} \\
\tilde c^\alpha &=&  c^\alpha - 2Q \phi^\alpha~, \ \ \  \alpha\not={\rm Killing}~.\label{alphaginv2}
\eea
These are all gauge invariant, with the exception of $\tilde A_\mu^i$ and $\tilde a_\mu^i$ for $i={\rm Killing}$, which transform as $\tilde A_\mu^i=\partial_\mu\xi^i$, $\delta\tilde a_\mu^i=-{2Q\over \sqrt{\lambda_i}}\partial_\mu\xi^i$.  One linear combination of these is gauge invariant and will yield a massive vector, the other transforms as a Maxwell field and will give a massless vector with a gauge invariant Maxwell action.

There is a curiosity that takes place when the internal manifold is flat, $R_{(N)}=0$.  As reviewed in Appendix \ref{appendixkilling}, this is the only case for which there can be harmonic forms which are also Killing, so that $\alpha={\rm Killing}$ is non-empty, and in this case the harmonic forms are precisely the Killing vectors.  As indicated in Eq.~\eqref{alphaginv1}, the gauge invariant combination is $A_\mu^\alpha-{1\over 2Q}\partial_\mu c^\alpha$, which combines the zero-modes of the flux field with the zero modes of the Kaluza-Klein photon coming from the graviton (the gravi-photon).  This combination will appear in $d$ dimensions as a massive vector, whose mass can be interpreted as arising from a kind of Anderson-Higgs mechanism; the gravi-photon eats the zero-mode of the form field to become massive.  This only happens for non-zero flux $Q^2>0$, so that by Eq.~\eqref{fluxreln} the $d$-dimensional space is negatively curved, $R_{(d)}<0$.  When $Q^2=0$, we see from  Eq.~\eqref{diffonfluxlowd} that the zero mode becomes gauge invariant and is no longer combined with the gravi-photon.  The scalar and the gravi-photon then become a separate massless vector and scalar.

\subsection{Reduced Action}

Using orthogonality of the various eigenfunctions to integrate out the internal dimensions, combining all the different terms and regrouping them into the combinations Eqs.~\eqref{gaugecombosstart}-\eqref{alphaginv2}, we find,
\begin{align}
\nn \frac{\mathcal{L}_0}{\sqrt{-g}} &= \varepsilon(h^0) + h^{\mu\nu,0}(\nabla_\mu \nabla_\nu \phi^0 - \Box \phi^0 g_{\mu\nu}) - \frac{R_{(d)}}{d}h^0 \phi^0\\
& + \frac{1}{2}\left(1-\frac{1}{N}\right)(\partial \phi^0)^2 -\left[\frac{1}{2}\left(1-\frac{2}{N}\right)\frac{R_{(d)}}{d} + \left(1-\frac{1}{N}\right)Q^2\right](\phi^0)^2~,\label{fluxaction0}\\
\nn\frac{\mathcal{L}_a}{\sqrt{-g}} &=\sum_{a\not={\rm conformal}} \varepsilon(\tilde h^a) - \frac{1}{2}\lambda_a\left( |\tilde h_{\mu\nu}^a|^2 -|\tilde h^a|^2\right)\\
\nn & + \frac{1}{2}\left\{\tilde h^{\mu\nu, a *}(\nabla_\mu \nabla_\nu F^a - \Box F^a g_{\mu\nu}) + \left[\left(1-\frac{1}{N}\right)\lambda_a - \frac{R_{(d)}}{d}\right] \tilde h^{a*} F^a + Q\lambda_a\tilde h^{a*}K^a +c.c.\right\}\\ 
\nn&+\frac{1}{2}\left(1-\frac{1}{N}\right)|\partial F^a|^2 + \frac{1}{2}\left(1-\frac{2}{N}\right)\left[\left(1-\frac{1}{N}\right)\lambda_a-\frac{R_{(d)}}{d}\right]|F^a|^2 - \left(1-\frac{1}{N}\right)Q^2 |F^a|^2\\
&-\frac{1}{2}\lambda_a|\partial K^a|^2-\frac{1}{2}\lambda_a^2 |K^a|^2 - \frac{1}{2}\lambda_a Q \left\{K^{a*} F^a+c.c.\right\}~\nn \\
&+\sum_{a={\rm conformal}} \varepsilon(\tilde h^a) - \frac{1}{2}\lambda_a \left(|\tilde h_{\mu\nu}^a|^2 -|\tilde h^a|^2\right) + \frac{1}{2}\left\{   Q\sqrt{\lambda_a}\tilde h^{a*}K^a      +c.c.\right\}-\frac{1}{2}|\partial K^a|^2-\frac{1}{2}\lambda_a |K^a|^2~,\label{fluxactiona}\\
\frac{\mathcal{L}_i}{\sqrt{-g}} &= \sum_i -\frac{1}{2}|\tilde F_{\mu\nu}^i|^2 - \frac{1}{4}|\tilde f_{\mu\nu}^i|^2 - \left(\lambda_i - \frac{2R_{(d)}}{d} - 2Q^2\right)|\tilde A_\mu^i|^2 - 2|Q \tilde A_\mu^i + \frac{1}{2}\sqrt{\lambda_i} \tilde a_\mu^i|^2~,\label{fluxi}\\
\frac{\mathcal{L}_{\alpha}}{\sqrt{-g}} &= \sum_{\alpha\not={\rm Killing}} -\frac{1}{2}|\tilde F_{\mu\nu}^\alpha|^2  + 2\left(  \frac{R_{(d)}}{d} + Q^2\right)|\tilde A_\mu^\alpha |^2 - 2|Q \tilde A_\mu^\alpha - \frac{1}{2} \partial_\mu \tilde c^\alpha|^2~\nn\\
& +\sum_{\alpha={\rm Killing}} -\frac{1}{2}|\tilde F_{\mu\nu}^\alpha|^2  +\frac{2R_{(d)}}{d} |\tilde A_\mu^\alpha |^2 ~,\label{fluxalpha}\\
\frac{\mathcal{L}_{\cal I}}{\sqrt{-g}} &=\sum_{\cal I} -\frac{1}{2}|\partial \phi^{\cal I}|^2 - \frac{1}{2}\left(\lambda_{\cal I} - \frac{2R_{(d)}}{d} - 2Q^2\right)|\phi^{\cal I}|^2~\label{fluxactionI},
\end{align}
where $\tilde F_{\mu\nu}^i$, $\tilde F_{\mu\nu}^\alpha$ are the standard Maxwell field strengths of the corresponding vectors.

In addition, we have terms that we have not displayed coming from the kinetic terms for the form fluctuations (the third line of Eq.~\eqref{quadratic}).  These are unaffected by the mixing with gravity, and look identical to Eq.~\eqref{pformstuk2} for $p=N-1$.  The only exceptions, which are affected and which we have included here, are the parts in Eqs.~\eqref{fluxgravdecomp1} and~\eqref{fluxgravdecomp2}; in particular, the kinetic term for these parts coming from the third line of Eq.~\eqref{quadratic} give the terms in Eqs.~\eqref{fluxactiona},~\eqref{fluxi} and~\eqref{fluxalpha} that are squares of the $p$-form components.

For the case $N=2$, there was the subtlety that the one-forms are mapped to themselves under the Hodge star, so that we had to split the index $\alpha$ into imaginary self-dual and imaginary anti-self dual parts, $\alpha_\pm$, as in Eq.~\eqref{imselfd}.  From the first term of Eq.~\eqref{mixingb}, we get the cross term $\sum_\alpha Qi\left(A_\mu^{\alpha^-}\partial^\mu c^{\alpha^+}-A_\mu^{\alpha^+}\partial^\mu c^{\alpha^-}\right)+c.c.$  After a field re-definition  $c^{\alpha^+}\rightarrow -ic^{\alpha^-}$, $c^{\alpha^-}\rightarrow ic^{\alpha^-}$, the $\alpha^+$ and $\alpha^-$ indices can be recombined into $\alpha$, giving the cross terms of Eq.~\eqref{fluxalpha} which are of the same form as $N>2$.

The part of the spectrum that can be immediately read off is $\mathcal{L}_{\cal I}$.  This gives 
a tower of massive scalars with mass $m_{\cal I}^2=\lambda_{\cal I} - \frac{2R_{(d)}}{d} - 2Q^2=\lambda_{\cal I}- \frac{2R_{(N)}}{N}$.  In terms of $R_{(N)}$ this is the same mass spectrum as for pure gravity.  In particular, the scalars are massless precisely when $\lambda_{\cal I}=\frac{2R_{(N)}}{N}$, i.e. for each modulus of the Einstein structure.
For the rest, we must diagonalize the action.

\subsection{Diagonalization}

We now diagonalize the various parts of the action that are mixed.  
The case $d=2$ is special due to the fact that gravitons have a vanishing kinetic term in this dimension, so we first restrict to the case $d\geq3$.  (As with the pure graviton, the case $d=1$ is trivial: $R_{(d)}=0$, every scalar indexed by $a$ is conformal, and everything that might need diagonalization vanishes or is set to zero by equations of motion.)

\textbf{\underline{Zero modes}:}
We start with the ${\cal L}_0$ sector in Eq.~\eqref{fluxaction0}.  We can diagonalize this graviton-scalar sector by the conformal transformation
\be h^0_{\mu\nu} = h^{'0}_{\mu\nu} -\frac{1}{d-2} \phi^0 g_{\mu\nu}~, \ee
giving 
\bea
\frac{\mathcal{L}_0}{\sqrt{-g}} &=& \varepsilon(h'^0) + \frac{d+N-2}{2N(d-2)}\left(-(\partial\phi^0)^2 + \frac{2R_{(d)}}{d}(\phi^0)^2\right) - \left(1-\frac{1}{N}\right)Q^2 (\phi^0)^2~.
\eea
This describes a massless graviton, and a non-ghost scalar with mass,
\be m^2_0=-{2R_{(d)}\over d}+{2(N-1)(d-2)\over d+N-2}Q^2~, \label{fluxzeromass}\ee 
Using Eq.~\eqref{fluxreln} we may write this as $m^2_0=-{2R_{(N)}\over N}+{2N(d-1)\over d+N-2}Q^2$, and we see that the volume modulus for positively curved ${\cal N} $ may be stabilized if the flux is sufficiently large.

\textbf{\underline{$a$ modes}:}
We next turn to the ${\cal L}_a$ sector in Eq.~\eqref{fluxactiona}.  

First consider the non-conformal sector. We can diagonalize by the transformation
\begin{align} 
\nn h^a_{\mu\nu}\rightarrow h^{'a}_{\mu\nu}&-\frac{d\left(1-\frac{1}{N}\right)\lambda_a-R_{(d)}}{d(d-1)\lambda_a-(d-2)R_{(d)}}F^a g_{\mu\nu}+ \frac{d\left(1+\frac{d-2}{N}\right)}{d(d-1)\lambda_a-(d-2)R_{(d)}}\partial_\mu\partial_\nu F^a \\
&-\frac{d Q \lambda_a}{d(d-1)\lambda_a-(d-2)R_{(d)}}K^a g_{\mu\nu}-\frac{d(d-2) Q}{d(d-1)\lambda_a-(d-2)R_{(d)}}\partial_\mu\partial_\nu K^a~, \ \ \ a\not={\rm conformal}~.
\end{align}
The Lichnerowicz bound Eq.~\eqref{lichbound} combined with the positivity of $Q^2$ in Eq.~(\ref{fluxreln}) implies that the denominator is strictly positive, so the transformation is always valid (except in the case $d=1$ where $R_{(d)}=0$, which we will treat separately). 

This decouples the graviton, leaving a Fierz-Pauli massive graviton of mass squared $\lambda_a$.  The remaining action for $F^a$ and $K^a$ is still kinetically mixed, and the mixing between the kinetic terms of the $F^a$ and $K^a$ scalars can be removed by the transformation,
\be F^a\rightarrow F^{'a} - \frac{d \,\lambda_a Q }{d\left(1-\frac{1}{N}\right)\lambda_a-R_{(d)}}K^a~, \ \ \ a\not={\rm conformal}~.\ee
Again the denominator is never zero.
The resulting mass matrix mixing the $F^a$ and $K^a$ scalars has eigenvalues
\begin{align} 
m^2_{a \pm} = \lambda_a+A \pm \sqrt{A^2+4\lambda_a B}~,\ \ a\neq\rm{conformal}~, \ \ {\rm where}\ \ \begin{cases} A\equiv \frac{(d-2)(N-1)Q^2}{d+N-2}-\frac{R_{(d)}}{d}~,\\ B\equiv \frac{(d-1)(N-1)Q^2}{d+N-2}~. \end{cases}
\label{amassmatrix}
\end{align}

Now consider the conformal sector. We can diagonalize by the transformation
\be h^a_{\mu\nu}\rightarrow h^{'a}_{\mu\nu} - \frac{d\,\sqrt{\lambda_a} Q}{d(d-1)\lambda_a-(d-2)R_{(d)}} K^a g_{\mu\nu} - \frac{d(d-2)\frac{Q}{\sqrt{\lambda_a}}}{d(d-1)\lambda_a-(d-2)R_{(d)}} \partial_\mu\partial_\nu K^a~, \ \ \ a={\rm conformal}~, \ee
which is valid since the denominator is always strictly positive by $\lambda_a=\frac{R_{(N)}}{N-1}$ for conformal scalars, combined with the positivity of $Q^2$ in Eq.~(\ref{fluxreln}).  This gives the final Lagrangian for the conformal part,
\begin{align} 
\nonumber \frac{\mathcal{L}_a}{\sqrt{-g}} \supset& \ \sum_{a=\rm{conformal}} \varepsilon(h^a) - \frac{1}{2}\lambda_a (|h_{\mu\nu}^a|^2 -|h^a|^2) \\
&+ \frac{d(d+N-2)\left(Q^2+\frac{R_{(d)}}{d}\right)}{(d+N-2)R_{(d)}+Nd(d-1)Q^2}\left[-\frac{1}{2}|\partial K^a|^2 - \frac{1}{2}\left(\lambda_a + \frac{2N(d-1)}{d+N-2}Q^2\right)|K^a|^2\right]~. 
\end{align}
The denominator of the kinetic term for the scalars is always strictly positive. Also, recall that $Q^2 + \frac{R_{(d)}}{d}=\frac{R_{(N)}}{N}$ by Eq.~(\ref{fluxreln}). Thus, there are no ghosts for a positively curved internal space, $R_{(N)}>0$. For a flat or negatively curved internal space, $R_{(N)}=0$ or $R_{(N)}<0$, no conformal scalars exist (see Appendix~\ref{appendixconformal}).

The masses of these scalars are
\be m_a^2 = \lambda_a + \frac{2N(d-1)}{d+N-2}Q^2~, \ \ a=\rm{conformal}~. \label{conformalmassflux}\ee
The mass is always strictly positive, which is seen by using $\lambda_a=\frac{R_{(N)}}{N-1}$ combined with the positivity of $Q^2$ in Eq.~(\ref{fluxreln}). 

\textbf{\underline{$i$ modes}:}
The Lagrangian ${\cal L}_i$, \eqref{fluxi}, describes two vectors for each $i$.
The mass matrix has eigenvalues,
\be m_{i\pm}^2 = \lambda_i -\frac{R_{(d)}}{d}\pm\sqrt{\left(\frac{R_{(d)}}{d}\right)^2+2Q^2\lambda_i}~. \label{iroots}\ee
The positive root is always greater than zero.  The negative is always greater than or equal to zero, and is equal to zero precisely when $i$ corresponds to a Killing vector.  Recall from Appendix~\ref{appendixkilling} that $\lambda_i\geq {2R_{(N)}\over N}$, with equality if and only if $i$ corresponds to a Killing vector.  In the case $R_{(N)}>0$, we may have such Killing vectors, and the negative root of Eq.~\eqref{iroots} will give one massless vector for each of the Killing vectors.  For the Killing vectors, the positive root remains massive, corresponding to the gauge invariant combination of the two vectors $\tilde A_\mu^i$ and $\tilde a_\mu^i$, and the negative root corresponds to a gauge non-invariant combination which transforms as a massless vector should.

In the case $R_{(N)}\leq 0$, we do not have Killing vectors among the $i$'s, because they don't exist when $R_{(N)}< 0$, and because they are harmonic and hence included among the $\alpha$'s when $R_{(N)}=0$.  The bound $\lambda_i\geq 0$ on the vector Laplacian ensures that the squared masses Eq.~\eqref{iroots} are not less than $0$.  

\textbf{\underline{$\alpha$ modes}:}
Finally, let us analyze the spectrum of the harmonic sector ${\cal L}_\alpha$ in Eq.~\eqref{fluxalpha}.  
For $\alpha\not={\rm Killing}$, we diagonalize by the transformation
\be \tilde A_\mu^\alpha \rightarrow \tilde A^{' \alpha}_\mu -\frac{d\, Q}{2R_{(d)}}\partial_\mu \tilde c^\alpha~, \ \ \ \alpha\not={\rm Killing}~.\label{transfal}\ee
Note that we do not have to worry about $R_{(d)}=0$ in the denominator of the transformation Eq.~\eqref{transfal}, because in this case Eq.~\eqref{fluxreln} tells us that $R_{(N)}>0$, which by the arguments of Appendix \ref{appendixkilling} implies that there are no harmonic vectors. We have,
\be \frac{\mathcal{L}}{\sqrt{-g}}=-\frac{1}{2}|\tilde F_{\mu\nu}^{'\alpha}|^2+\frac{2R_{(d)}}{d}|A_\mu^{'\alpha}|^2- \frac{1}{2}\left(1+\frac{d }{R_{(d)}}Q^2\right) |\partial_\mu \tilde c^\alpha|^2~, \ \ \ \alpha\not={\rm Killing}~. \ee
We find massless scalars when $R_{(N)}>0$, which are never ghostly since $1+\frac{d }{R_{(d)}}Q^2>0$ when $R_{(N)}> 0$, and vectors of mass squared $m^2=-{2R_{(d)}\over d}=2Q^2-{2R_{(N)}\over N}$.  This mass squared is always greater than zero, since harmonic vectors do not exist on positively curved manifolds. 

The case $R_{(N)}=0$ is the case when any harmonic vectors are also Killing, so we do not need to diagonalize in this case, and we have only massive vectors of mass $m^2=-{2R_{(d)}\over d}$, which is always positive because by Eq.~\eqref{fluxreln}  $R_{(d)}<0$ when $R_{(N)}=0$ and $Q^2>0$.

\noindent {\bf \underline{$d=2$ case}:}\\
In $d=2$, gravitons are non-dynamical, so we must solve for the gravitons as auxiliary fields through their equations of motion. 
The kinetic part of the graviton, $\sqrt{-g}\epsilon(h)$ with $\epsilon(h)$ as in Eq.~\eqref{epsilondef}, is a total derivative.  Dropping this total derivative, the zero mode sector after reduction has the Lagrangian
\begin{align}
\nn \frac{\mathcal{L}_0}{\sqrt{-g}} &= h^{\mu\nu,0}(\nabla_\mu \nabla_\nu \phi^0 - \Box \phi^0 g_{\mu\nu}) - \frac{R_{(2)}}{2}h^0 \phi^0\\
& + \frac{1}{2}\left(1-\frac{1}{N}\right)(\partial \phi^0)^2 -\left[\frac{1}{2}\left(1-\frac{2}{N}\right)\frac{R_{(2)}}{2} + \left(1-\frac{1}{N}\right)Q^2\right](\phi^0)^2~.\
\end{align}
This is just a linearization of a version of dilaton gravity in two dimensions~\cite{Grumiller:2002nm}.

The part of the reduced action corresponding to scalar eigenfunctions of the Laplacian becomes
\begin{align}
\nn\frac{\mathcal{L}_a}{\sqrt{-g}} &=\sum_{a\not={\rm conformal}}  - \frac{1}{2}\lambda_a |\tilde h_{\mu\nu}^a|^2 + \frac{1}{2}\lambda_a|\tilde h^a|^2\\
\nn & + \frac{1}{2}\left\{\tilde h^{\mu\nu, a *}(\nabla_\mu \nabla_\nu F^a - \Box F^a g_{\mu\nu}) + \left[\left(1-\frac{1}{N}\right)\lambda_a - \frac{R_{(2)}}{2}\right] \tilde h^{a*} F^a + Q\lambda_a\tilde h^{a*}K^a +c.c.\right\}\\ 
\nn&+\frac{1}{2}\left(1-\frac{1}{N}\right)|\partial F^a|^2 + \frac{1}{2}\left(1-\frac{2}{N}\right)\left[\left(1-\frac{1}{N}\right)\lambda_a-\frac{R_{(2)}}{2}\right]|F^a|^2 - \left(1-\frac{1}{N}\right)Q^2 |F^a|^2\\
&-\frac{1}{2}\lambda_a|\partial K^a|^2-\frac{1}{2}\lambda_a^2 |K^a|^2 - \frac{1}{2}\lambda_a Q \left\{K^{a*} F^a+c.c.\right\}~\nn \\
&+\sum_{a={\rm conformal}}- \frac{1}{2}\lambda_a |\tilde h_{\mu\nu}^a|^2 + \frac{1}{2}\lambda_a|\tilde h^a|^2 + \frac{1}{2}\left\{   Q\sqrt{\lambda_a}\tilde h^{a*}K^a      +c.c.\right\}-\frac{1}{2}|\partial K^a|^2-\frac{1}{2}\lambda_a |K^a|^2~.
\end{align}

For the non-conformal part, varying with respect to $h^{\mu\nu,a}$ gives us an expression for the non-dynamical graviton,
\be h_{\mu\nu}^a = \frac{1}{\lambda_a}\nabla_\mu\nabla_\nu F^a -\frac{1}{\lambda_a}\left[\left(1-\frac{1}{N}\right)\lambda_a-\frac{R_{(2)}}{2}\right]F^a g_{\mu\nu}+ Q K^a g_{\mu\nu}~, \ \ \ a\not={\rm conformal}~.\ee
Substituting this back into the action, we find
\begin{align} 
\nn\frac{\mathcal{L}_a}{\sqrt{-g}}\supset&\sum_{a\not={\rm conformal}} -\frac{1}{2}\left[\left(1-\frac{1}{N}\right)-\frac{R_{(2)}}{2\lambda_a}\right]|\partial F^a|^2 -\frac{1}{2}\left(1-\frac{R_{(2)}}{\lambda_a}\right)\left[\left(1-\frac{1}{N}\right)\lambda_a-\frac{R_{(2)}}{2}\right]|F^a|^2 \\ \nn 
& -\left(1-\frac{1}{N}\right)Q^2|F^a|^2
\nn -\frac{1}{2}\lambda_a (\partial K^a)^2 - \frac{1}{2}\left(\lambda_a+2Q^2\right)\lambda_a |K^a|^2\\
&+{1\over 2}\left\{ \left[\left(-3+\frac{2}{N}\right)\lambda_a + R_{(2)}\right] Q {K^a}^\ast F^a -Q\partial_\mu {F^a}^\ast \partial^\mu K^a +c.c. \right\} ~. 
\end{align}
The kinetic cross-terms can be eliminated by the transformation,
\be F^a\rightarrow {F^a}^{'} - \frac{Q \lambda_a}{\left(1-\frac{1}{N}\right)\lambda_a-\frac{R_{(2)}}{2}}K^a~,\ \ \ a\not={\rm conformal}~. \ee  As with the previous cases, the denominator is never zero.
After canonically normalizing the kinetic terms, the resulting mass matrix has eigenvalues
\be m_{\pm}^2=\lambda_a-\frac{R_{(2)}}{2}\pm\sqrt{\frac{R_{(2)}^2}{4}+4\left(1-\frac{1}{N}\right)Q^2\lambda_a}~,\ \ \ a\not={\rm conformal}~.  \label{2dmass}\ee
This is exactly the same as the earlier formula Eq.~(\ref{amassmatrix}) restricted to $d=2$.

For the conformal sector, the equations of motion give
\be h_{\mu\nu}^a = \frac{Q}{\sqrt{\lambda_a}} K^a g_{\mu\nu}~,\ \ \ a={\rm conformal}~.  \ee
Substituting back into the action gives
\be \frac{\mathcal{L}_a}{\sqrt{-g}}\supset\sum_{a={\rm conformal}}-\frac{1}{2}|\partial K^a|^2 - \frac{1}{2}\left(\lambda_a+2Q^2\right) |K^a|^2~,\ \ \ a={\rm conformal}~.  \ee

\noindent {\bf \underline{$d=1$ case}}:\\  
In $d=1$ we have $R_{(d)}=0$.  All the graviton kinetic terms, Fierz-Pauli mass terms, and vector kinetic terms vanish identically.  All that remains of ${\cal L}_0$ is the volume modulus $\phi^0$ with mass $m^2=2\left(1-1/N\right)Q^2$.  All that remains of the graviton in ${\cal L}_a$ for the non-conformal modes are cross terms $\sim h F,hK$, so $h^a$ is a multiplier which sets $K^a=-{N-1\over NQ}F^a$. Plugging back in, we find $F^a$ remains as a free scalar with mass $m^2=\lambda_a-2Q^2$. In the case of the conformal modes, all that remains of the graviton is a cross term $\sim hK$, so $h^a$ is a multiplier which sets $K^a=0$.  Nothing dynamical remains of the vectors in ${\cal L}_i$, since the vector kinetic terms vanish.  ${\cal L}_\alpha$ is empty because $R_{(d)}=0$ means $R_{(N)}>0$ so there are no harmonic vectors.  Thus the spectrum for $d=1$ is just the zero mode, the non-conformal scalars of ${\cal L}_a$, and the scalars of ${\cal L}_{\cal I}$.

\subsection{Spectrum\label{specfluxsec}}

In summary, the spectrum is 
\begin{itemize}
\item One massless graviton (dilaton gravity in $d=2$),
\item One zero mode scalar for the volume modulus with a curvature-dependent mass $m^2_0=-{2R_{(N)}\over N}+{2N(d-1)\over d+N-2}Q^2$ (part of dilaton gravity in $d=2$), 
\item A tower of Fierz-Pauli massive gravitons, one for each eigenvector of the scalar Laplacian with eigenvalue $\lambda_a>0$, with mass $m_a^2=\lambda_a$,
\item A tower of massive scalars, two for each non-conformal eigenvalue of the scalar Laplacian labeled by $a$, with masses given by Eq.~\eqref{amassmatrix} (one of the pair is not dynamical for $d=1$), 
\item A tower of massive scalars for each conformal scalar when $N\geq 2$, with masses given by $m_a^2 = \lambda_a + \frac{2N(d-1)}{d+N-2}Q^2$ where $\lambda_a=\frac{R_{(N)}}{N-1}$ (not dynamical for $d=1$),
\item A tower of vectors for each co-exact form labeled by $i$, with masses given by Eq.~\eqref{iroots}.   One is massless for each Killing vector,
\item Vectors with mass $m_\alpha^2=-\frac{2R_{(d)}}{d}$ for each harmonic form,
\item A massless scalar for each non-Killing harmonic form labeled by $\alpha$,
\item A tower of massive scalars for the transverse traceless tensor modes of the Lichnerowicz operator, in general massive with mass $m_{\cal I}^2=\lambda_{\cal I}- \frac{2R_{(N)}}{N}$. They are massless when $\lambda_{\cal I}=\frac{2R_{(N)}}{N}$, i.e. for each modulus of the Einstein structure. 
\end{itemize} 
In addition, we have the remaining parts of the $(N-1)$-form field decomposition which do not mix with the graviton.

Many of the ingredients are non-dynamical for low dimensions.  For instance, massless gravitons are non-dynamical for $d\leq 3$, the massive gravitons and massless vectors are non-dynamical in $d\leq 2$, and the vector fields are non-dynamical for $d=1$.

Let us comment on the case $N=1$, where ${\cal N} $ is the circle and the extra dimensional metric has only one component $\gamma_{yy}=1$.  There are no co-exact vectors other than the single Killing vector which is constant around the circle.  This is also the only harmonic one-form.  There are no transverse traceless tensors, and no non-conformal scalars (every vector is a conformal Killing vector for $N=1$).  The scalar eigenfunctions can be chosen to be simply the Fourier modes, $\psi_a={1\over \sqrt{2\pi {\cal R}}}e^{iay/ {\cal R}}$, where ${\cal R}$ is the radius of the circle and $a$ ranges over all the integers, with $a=0$ the zero mode.  The eigenvalues are $\lambda_a=a^2/ {\cal R}^2$.   In this case, the graviton ansatz Eq.~\eqref{ansatzH2} simplifies to the simple Fourier expansion Eq.~\eqref{ansatzHn1}.  The $p$-form field is simply a scalar, $A$, with a background value $\partial_y A=Q$.  The $d$-dimensional space is negatively curved, with curvature $R_{(d)}=-Q^2 d$.  The expansion Eq.~\eqref{fluxdecomp} of the fluctuation of $A$ becomes a simple Fourier transform,
\be \tilde A=\frac{1}{\sqrt{2\pi \mathcal{R}}}\left[c^0+\sum_{a=1}^\infty\left(    a^ae^{iay/ {\cal R}}+c.c.\right)\right].\label{n1relphi}\ee
As for the duality relations, Eq.~\eqref{hodgelowrelations2} is empty, and Eq.~\eqref{hodgelowrelations1} relates the space of scalar eigenfunctions to itself and for our choice of basis must be supplemented by an additional factor of $i$ to remain true, $ \psi^a = \frac{i}{\sqrt{\lambda_{a}}}\ast d\psi_{a}$.  Then Eq.~\eqref{tildesaeq} remains true and reads just like Eq.~\eqref{n1relphi} if we redefine $a^a\rightarrow -i a^a$.

In this case, we recover a massless graviton $h_{\mu\nu}^0$, a massive vector $A_\mu^0$ with mass squared $-2R_{(d)}/d$, the massive dilaton $\phi^0$ with mass squared $-2R_{(d)}/d$, a tower of massive graviton doublets $h_{\mu\nu}^a$ with longitudinal modes $A_\mu^a$, $\bar \phi^a$, and a tower of scalars $K^a$.  As mentioned at the end of Section \ref{fluxgaugecombsec}, the presence of flux can be interpreted as an Anderson-Higgs mechanism: the zero mode of the zero-form field is eaten by the Kaluza-Klein vector, causing it to become massive. 

\subsection{Stability}

\textbf{\underline{Gravitons}:} 
The graviton spectrum is identical to the case of pure gravity with no flux.  For gravitons on a flat or negatively curved space, stability occurs only when the Fierz-Pauli mass term is negative, which  never happens.  For positively curved spaces, gravitons are stable as long as their masses are above the Higuchi bound \cite{Higuchi:1986py},
\be m^2\geq{d-2\over d(d-1)}R_{(d)}~.\ee

Since the mass squared of the graviton is just the eigenvalue, $m^2=\lambda_a$, to violate or saturate the Higuchi bound we would need to have 
\be \lambda_a \leq {d-2\over d-1}{R_{(d)}\over d}={d-2\over d-1}\left({R_{(N)}\over N}-Q^2\right)\leq {d-2\over d-1}{R_{(N)}\over N}~.\ee
This is inconsistent with the Lichnerowicz bound Eq.~\eqref{lichbound}, $\lambda_a \geq {R_{(N)}\over N-1}.$  Thus the gravitons are always stable, the Higuchi bound is never saturated, and there is never a partially massless graviton in the spectrum.

\textbf{\underline{Vectors}:}  Vectors are unstable only when their mass squared is negative, which never occurs, so no instability arises from the vectors.

\textbf{\underline{Scalars}:} Any instabilities always arise from the scalar modes, which may potentially become tachyonic.

\begin{itemize}
\item The volume modulus $\phi^0$ is always stable for $R_{(d)} \leq 0$.  For a positively curved internal manifold, $R_{(d)} > 0$, there is  a window of stability,
 \be {2N(d-1)\over d+N-2}Q^2\geq {2R_{(N)}\over N}~.\ee Recall that this sector is always unstable for positive curvature in the pure graviton case, $Q^2=0$, so the flux has a stabilizing effect.  In the case $d=2$, where we have dilaton gravity, there are no local degrees of freedom \cite{Kim:1994nq}, so we may call this stable in all cases if we are only concerned with local degrees of freedom.

\item There may be instabilities in the $\phi^{\cal I}$, the sector corresponding to transverse traceless tensor modes of the Lichnerowicz operator.  In terms of the curvature of the internal manifold, the condition for stability is identical to that of the pure gravity case: the mass squared is given by $m^2_{\cal I}=\lambda_{\cal I} - \frac{2R_{(N)}}{N} $. These are never unstable in the case of compactifications on spheres or other quotients of maximally symmetric spaces. 

\item Additional instabilities can occur in the scalar sector that corresponds to scalar eigenmodes of the Laplacian. 
The regime of stability is governed by Eq.~\eqref{amassmatrix}. In the following subsection, we will look more closely at this spectrum in a specific and familiar setup, namely for compactifications on spheres.
\end{itemize}

\subsection{Spheres}

While the spectrum of scalars in the scalar eigenmode sector is complicated in general, it is simple to make contact with known stability regimes of standard Freund-Rubin compactifications when the case where the internal manifold is a sphere (see e.g.~\cite{Bousso:2002fi}). Here we summarize some results from these formulae in the case of spheres for $d,N\geq 2$.  The scalar eigenmodes are $\lambda_a = \frac{l(l+N-1)}{\mathcal{R}^2}$, where $l=0,1,2,\cdots$ and $\mathcal{R}$ is the radius of the sphere and is related to the curvature by $R_{(N)}=\frac{N(N-1)}{\mathcal{R}^2}$.  The $l=0$ mode is the zero modes, and its stability is governed by Eq.~\eqref{fluxzeromass}.  The $l=1$ modes are the conformal scalars, with mass squared given by Eq.~\eqref{conformalmassflux}, and they are always stable.   The $l\geq 2$ modes are the non-conformal scalars, and their stability is governed by the lesser of Eq.~\eqref{amassmatrix}.  The $l=0$ and $l\geq2$ modes are the potential sources of instability.  We have:\\

\underline{$d>2$}:
\begin{itemize}
\item \underline{$R_{(d)}<0$}: The $l=0$ modes are always stable. For the $l\geq2$ modes, the negative solution must be compared to the Breitenlohner-Friedmann bound Eq.~\eqref{BFboundref} to assess stability. We have,
\begin{itemize}
\item $N=2,3$: Stable,
\item $N$ even, $N\geq 4$: Unstable to one or more $l\leq \frac{N}{2}$ modes when $\Lambda$ is above some critical value which is $>0$. 
Stable when $\Lambda$ is below this critical value, in particular, stable when $\Lambda\leq 0$.
\item $N$ odd, $N\geq 5$: Entire $\Lambda > 0$ regime is unstable to one or more $l\leq \frac{N-1}{2}$ modes.  $\Lambda=0$ is stable, with $l=\frac{N-1}{2}$ exactly saturating the Breitenlohner-Friedmann bound.  $\Lambda<0$ is stable.
\end{itemize}
\item \underline{$R_{(d)}>0$}: These solutions always have $\Lambda>0$. The $l=0$ modes are now stable in the range ${2(d-2)\over d-1}\Lambda\geq {d+N-2\over d}R_{(d)}$, and so are unstable below some critical value for $\Lambda$. 
For the $l\geq2$ modes, we have
\begin{itemize}
\item $N=2,3$:  
Stable. 
\item $N=4$:  
Unstable to the $l=2$ mode above a critical value of $\Lambda$.   There is a window of stability between this critical value and the critical value for the zero mode.
\item $N\geq5$: Unstable to one or more $l\geq 2$ modes above a critical value of $\Lambda$, with no window of stability between this critical value and the critical value for the zero mode.
\end{itemize}
\item \underline{$R_{(d)}=0$}: These solution satisfy $\Lambda={1\over 2}\left(1-{1\over N}\right)R_{(N)}>0$. The $l=0$ modes are always stable.  For the higher modes we have,
\begin{itemize}
\item $N=2,3$: Stable,
\item $N\geq 4$: Unstable to one or more $l\geq2$ modes.\\
\end{itemize}
\end{itemize}

In $d=2$, the decoupled zero mode with mass Eq.~\eqref{fluxzeromass} is replaced by a dilation gravity theory, which carries no local degrees of freedom, so there is no local instability here~\cite{Kim:1994nq}. Thus, any instabilities come from only for the $l\geq2$ modes. Apart from this, the main differences occur in the case $N=4$. We have,\\

\underline{$d=2$}: 
\begin{itemize}
\item \underline{$R_{(d)}<0$}: An additional window of stability opens for $N=4$ above a new critical $\Lambda>0$. Otherwise the same as $d>2$. 
\item \underline{$R_{(d)}>0$}:  We have,
\begin{itemize}
\item $N=2,3,4$: Stable,
\item $N\geq 5$:  Unstable to one or more $l\geq 2$ modes when $\Lambda$ is larger than some critical value. 
\end{itemize}
\item \underline{$R_{(d)}=0$}: We have,
\begin{itemize}
\item $N=2,3,4$: Stable,
\item $N\geq 5$: Unstable to one or more $l\geq2$ modes. 
\end{itemize}
\end{itemize}

\section{Discussion}

We have derived the full Kaluza-Klein tower of fluctuations for scalars, vectors, $p$-forms, linearized gravity and flux compactifications, on a general smooth manifold of arbitrary dimension, using Hodge decomposition theorems.  We have performed these calculations in a fully gauge invariant way, at the level of the action. 

The dimensional reduction of linear theories has been studied in the past for large classes of compactifications, but we feel the calculations presented here, based on the action and performed in a gauge invariant manner, are more streamlined than those given previously based on the equations of motion and/or restricted to specific gauges.  

For instance, we can see at the level of the action the instabilities occurring in~\cite{Bousso:2002fi, Frolov:2003yi, Contaldi:2004hr, Martin:2004wp}.  They are a general feature of Kaluza-Klein reductions; for example, any reduction of gravity on a positively curved spacetime will feature instabilities coming from the volume modulus. 

We have worked with only linearized theories, so our calculations ignore any effects due to non-linear interactions. 
However, another virtue of our method is that it is straightforward, if tedious, to extend to interactions.  One simply plugs the ans\"atze we have given for the fields into the full non-linear action.  The interaction terms mixing the modes in $d$ dimensions will then have coefficients containing triple and higher overlap integrals involving the various pieces of the Hodge decompositions.

\bigskip
{\bf Acknowledgments:}
We would like to thank James Bonifacio, Adam Brown, Alex Dahlen, Ori Ganor, Victor Ivo, Matt Johnson, Dan Lee and Robert Lipshitz for useful discussions and comments.  Research at Perimeter Institute is supported by the Government of Canada through Industry Canada and by the Province of Ontario through the Ministry of Research and Innovation. The 
work of KH was made possible in part through the support of a grant from the John Templeton 
Foundation (the opinions expressed in this publication are those of the authors and do not necessarily 
reflect the views of the John Templeton Foundation).  The work of CZ is supported by the Berkeley Center for Theoretical Physics, by the National Science Foundation (award numbers 0855653 and 0756174), by fqxi grant RFP3-1004, by the U.S. Department of Energy under Contract DE-AC02-05CH11231, and by an NSF Graduate Fellowship.  The work of JL is supported by NSF grant AST-0908365, a KITP Scholarship under Grant no. NSF PHY05-51164, and a Guggenheim Fellowship.  KH and CZ would like to thank Kavli IPMU for hospitality while part of this work was completed. 

\appendix

\section{Hodge Eigenvalue Decomposition for Forms}\label{appendixforms}

In this Appendix, we review some facts about the mathematics of $p$-forms and the Hodge decomposition, which we will need to decompose the higher dimensional fields in terms of components that are convenient for the Kaluza-Klein reduction.  For a more extensive introduction, see for example \cite{nakahara2003geometry}.

\subsection{Review of $p$-forms}

For (in general complex-valued) $p$-forms $\omega$ and $\eta$ on an $N$-dimensional manifold ${\cal N}$ with metric $\gamma$ and covariant derivative $\nabla$, we define the positive definite inner product
\be (\omega,\eta)=\int d^Ny\sqrt{\gamma}{1\over p!}\omega_{n_1\cdots n_p}^\ast\eta^{ n_1\cdots n_p}~.\label{formprod}\ee

The standard exterior derivative taking $p$-forms to $(p+1)$-forms is, 
\be (d\omega)_{n_1\ldots n_{p+1}}=(p+1)\nabla_{[n_1}\omega_{n_2\ldots n_{p+1}]}~.\ee
A $p$-form $\omega$ is {closed} if $d\omega=0$, and the space of closed $p$-form fields is denoted $\Lambda_{\rm closed}^p({\cal N})$.
A $p$-form $\omega$ is {exact} if $\omega=d\eta$, for some $(p-1)$-form $\eta$. The space of exact $p$-form fields is denoted by $\Lambda_{\rm exact}^p({\cal N})$.  An exact form is closed, since $d^2=0$, so $\Lambda_{\rm exact}^p({\cal N})$ is a subspace of $\Lambda_{\rm closed}^p({\cal N})$.  

The co-exterior derivative taking $p$-forms to ($p-1$)-forms is defined by
\be (d^\dag\omega)_{ n_1\ldots n_{p-1}}=-\nabla^m\omega_{m n_1\ldots n_{p-1}}~.\label{coextd}\ee
The operator $d^\dag$ is the adjoint of $d$ with respect to the inner product Eq.~\eqref{formprod}, i.e.
$(d\alpha,\omega)=(\alpha,d^\dag\omega)$, for $p$-form $\omega$ and $(p-1)$-form $\alpha$.
A $p$-form $\omega$ is {co-closed} if $d^\dagger\omega=0$, and the space of co-closed $p$-form fields is denoted $\Lambda_{\rm co-closed}^p({\cal N})$.
A $p$-form $\omega$ is {co-exact} if $\omega=d^\dagger\eta$, for some $(p+1)$-form $\eta$. The space of co-exact $p$-form fields is denoted by $\Lambda_{\rm co-exact}^p({\cal N})$.  A co-exact form is co-closed, since $(d^\dag)^2=0$, so $\Lambda_{\rm co-exact}^p({\cal N})$ is a subspace of $\Lambda_{\rm co-closed}^p({\cal N})$.  

The Hodge Laplacian is defined by
\begin{equation}
\Delta=(d+d^\dag)^2=dd^\dag+d^\dag d~,
\end{equation}
or in terms of components for a $p$-form $\omega$,
\bea &&(\Delta\omega)_{ n_1\ldots  n_p} = -\nabla^2\omega_{ n_1\ldots  n_p}+\sum_{i=1}^p\left[\nabla^ m,\nabla_{ n_i}\right]\omega_{ n_1,\ldots, n_{i-1}, m, n_{i+1}\ldots n_{p}} \nn\\
&& = -\nabla^2\omega_{ n_1\ldots  n_p}+\sum_{i=1}^pR^ m_{\  n_i}\omega_{ n_1,\ldots, n_{i-1}, m, n_{i+1}\ldots n_{p}}-2\sum_{i,j=1\ (i<j)}^pR^{ m_1\ m_2}_{\ \ \ n_i \ \  n_j}\omega_{ n_1,\ldots, n_{i-1}, m_1, n_{i+1}\ldots n_{j-1},m_2, n_{j+1}\ldots n_{p}}~, \nn\\\label{Hodgelapgen}
\eea
where $R^m_{\ npq}$ is the Riemann curvature.  A $p$-form $\omega$ is called {harmonic} if $\Delta\omega=0$.  A form is harmonic if and only if it is closed and co-closed.  The vector space of harmonic $p$-forms is denoted $\Lambda_{\rm harm}^p({\cal N})$.

The Laplacian is self-adjoint in the scalar product Eq.~\eqref{formprod},
\be (\omega,\Delta\eta)=(\Delta\omega,\eta)~.\ee
We can decompose the space of $p$-forms into eigenspaces of the Laplacian,
\be \label{eigenvaluedecomposition} \Lambda^p({\cal N})=\sum_{\oplus\lambda}E^{p}_\lambda({\cal N})~.\ee
where $E_\lambda^p({\cal N})$ are the subspaces $\{\omega\in \Lambda^p({\cal N})|\Delta\omega=\lambda\omega\}$.  Each subspace is finite dimensional.  We only consider those $\lambda$'s such that the subspaces are non-trivial, and this forms the spectrum of the Laplacian.  Due to self-adjointness and non-negativity of the Laplacian, the spectrum is real and non-negative,  
$\lambda\geq 0$, and the eigenspaces for different $\lambda$ are orthogonal with respect to the inner product \eqref{formprod}.  
The zero eigenspace is the same as the space of harmonic forms $E^{p}_{\lambda=0}({\cal N})=\Lambda^p_{\rm harm}({\cal N})$.

We have the Hodge decomposition theorem, giving the following orthogonal (under the inner product Eq.~\eqref{formprod}) direct sum decomposition
\be\label{hodgedecomposition} \Lambda^p({\cal N})=\Lambda^p_{\rm exact}({\cal N})\oplus \Lambda^p_{\rm co-exact}({\cal N})\oplus \Lambda_{\rm harm}^p({\cal N})~.\ee
This means that any $p$-form $\omega$ can be written uniquely as a sum of an exact form, a co-exact form, and a harmonic form.
We also have $\Lambda^p_{\rm closed}({\cal N})= \Lambda^p_{\rm exact}({\cal N})\oplus\Lambda_{\rm harm}^p({\cal N}), $ and
 $\Lambda^p_{\rm co-closed}({\cal N})= \Lambda^p_{\rm co-exact}({\cal N})\oplus\Lambda_{\rm harm}^p({\cal N}). $  The maps 
 \be \Lambda^p_{\rm co-exact}({\cal N})\overset{d}{\rightarrow} \Lambda^{p+1}_{\rm exact}({\cal N}),  \ \ \ \Lambda^p_{\rm co-exact}({\cal N})\overset{d^\dag}{\leftarrow} \Lambda^{p+1}_{\rm exact}({\cal N})~,\label{bijectionddag}\ee
 are bijections.

The differential and co-differential commute with the Laplacian,
\be d\Delta=\Delta d~,\ \ \  d^\dag\Delta=\Delta d^\dag~.\ee
Thus $\Lambda^p_{\rm exact}({\cal N})$ and $\Lambda^p_{\rm co-exact}({\cal N})$ are invariant subspaces of $\Delta$. 
In fact, The Hodge decomposition commutes with the eigenvalue decomposition.  Thus we can do a Hodge decomposition on each eigenspace separately for $\lambda\not=0$ (the $\lambda=0$ case is already done because $E_0^p({\cal N})=\Lambda^p_{\rm harm}({\cal N})$),
\be E_\lambda^p({\cal N})=E_{\lambda,{\rm exact}}^p({\cal N})\oplus E_{\lambda,{\rm co-exact}}^p({\cal N})~,\ee
and the spaces $\Lambda^p_{\rm exact}({\cal N})$ and $\Lambda^p_{\rm co-exact}({\cal N})$ can be decomposed into eigenspaces
\bea \Lambda^p_{\rm exact}({\cal N})&=&\sum_{\oplus\lambda\not=0}E^{p}_{\lambda,{\rm exact}}({\cal N})~, \\ \Lambda^p_{\rm co-exact}({\cal N})&=&\sum_{\oplus\lambda\not=0}E^{p}_{\lambda,{\rm co-exact}}({\cal N})~.\eea
The following sequences are exact for $\lambda>0$, 
 \bea &&\cdots\overset{d}{\rightarrow} E^p_{\lambda}({\cal N})\overset{d}{\rightarrow}E^{p+1}_\lambda({\cal N})\overset{d}{\rightarrow}\cdots \\  &&\cdots\overset{d^\dag}{\leftarrow} E^p_{\lambda}({\cal N})\overset{d^\dag}{\leftarrow}E^{p+1}_\lambda({\cal N})\overset{d^\dag}{\leftarrow}\cdots \eea
The kernel of $d$ on $E^p_{\lambda}({\cal N})$ is $E^p_{\lambda,{\rm exact}}({\cal N})$, and the kernel of $d^\dag$ on $E^p_{\lambda}({\cal N})$ is $E^p_{\lambda,{\rm co-exact}}({\cal N})$.

\subsection{The Kaluza-Klein Ansatz\label{KKanssection}}

The result of all this is that any $p$-form $\omega$ can be written as
\be \label{formdecompg} \omega_{n_1\cdots n_p}=\sum_{i_p} a^{i_p}Y_{i_p,n_1\cdots n_p}+\sum_{i_{p-1}} b^{i_{p-1}}\(dY_{i_{p-1}}\)_{n_1\cdots n_p}+\sum_{\alpha_p}c^{\alpha_p} Y_{\alpha_p,n_1\cdots n_p}~.
\ee
Here, $Y_{i_p}$ is a basis of co-closed $p$-forms, 
\be \nabla^m Y_{i_p,m n_2\cdots n_p}=0~,\ee
which are also eigenvalues of the Laplacian with positive eigenvalues
\be \Delta Y_{i_p,n_1\cdots n_p}=\lambda_{i_p}Y_{i_p,n_1\cdots n_p}~,\ \ \ \lambda_{i_p}>0~.\ee
 The index $i_p$ runs over the basis, including multiplicities for the eigenspaces.
The $Y_{\alpha_p}$ are a basis of harmonic $p$-forms,
\be \Delta Y_{\alpha_p,n_1\cdots n_p}=0~,\ee
(which, like all harmonic forms, are closed and co-closed, $ \nabla^m Y_{\alpha_p,mn_2\cdots n_p}=\nabla_{[n_1}Y_{\alpha_p,n_2\cdots n_{p+1}]}=0$).
The $a^{i_p}$'s, $b^{i_{p-1}}$'s and $c^{\alpha_p}$ in Eq.~\eqref{formdecompg} are constant coefficients which are unique for a given form $\omega$. 

 We also choose the bases above to be orthonormal in the product Eq.~\eqref{formprod}.  For the co-exact forms we have
 \be {1\over p!}\int d^{N}y \sqrt{\gamma}\ Y_{i_p,n_1\cdots n_p}^\ast  Y_{j_p}^{n_1\cdots n_p} = \delta_{i_p j_p}~,\ee
and for the harmonic forms we have
\be {1\over p!}\int d^{N}y \sqrt{\gamma}\ Y_{\alpha_p,n_1\cdots n_p}^\ast  Y_{\beta_p}^{n_1\cdots n_p} = \delta_{\alpha_p \beta_p}~.\label{harmonicnorma}\ee
The three parts of the decomposition Eq.~\eqref{formdecompg} are all orthogonal to each other.
The expression Eq.~\eqref{formdecompg} is an expansion of a $p$-form over a complete orthonormal basis of forms, a generalization of a Fourier expansion, one which is well suited for Kaluza-Klein reductions.  

Finally, we will impose that the bases have the following properties under complex conjugation,
\be Y_{i_p,n_1\cdots n_p}^\ast=\eta_{i_p} \, Y_{\bar i_p,n_1\cdots n_p}~,\ \ \ Y_{\alpha_p,n_1\cdots n_p}^\ast=\eta_{\alpha_p}\,Y_{\bar\alpha_p,n_1\cdots n_p}~.
\ee
Here $\bar i_p$ and $\bar \alpha_p$ represent some involution on the set of indices $i_p$, $\alpha_p$ respectively, and $\eta_{i_p}$, $\eta_{\alpha_p}$ are possible phases.\footnote{As an example, if we were considering the space of scalar functions on the two sphere, we could use the basis of spherical harmonics $Y_{l,m}$, indexed by the pair $l,m$ with $l=0,1,2,\cdots $ and $m=-l,-l+1\cdots,l$.  The case $l=0$ is the harmonic zero form and is the only value of the index $\alpha_0$.  The index $i_0$ then stands for the rest of the $l$'s and $m$'s collectively.  Under complex conjugation, the usual spherical harmonics transform as $Y_{l,m}^\ast=(-1)^mY_{l,-m}$, so the involution on the indices in this case is $\overline{(l,m)}=(l,-m)$ and the phase is $\eta_{l,m}=(-1)^m$.}  If the form $\omega$ which we are expanding in Eq.~\eqref{formdecompg} happens to be real (as is the case in our Kaluza-Klein decompositions where the higher dimensional fields are real), then the coefficients will satisfy a reality condition,
\be  a^{i_p\ast}= (\eta_{i_p}^{-1}) a^{\bar i_p}~,\ \ \ b^{i_{p-1}\ast}=(\eta_{i_{p-1}}^{-1}) b^{\bar i_{p-1}}~,\ \ \ c^{\alpha_p\ast}=(\eta_{\alpha_p}^{-1})c^{\bar\alpha_p}~.\ee

For the case $p=0$, where we are considering scalar functions $\phi$ on the manifold ${\cal N}$, there is always only one harmonic function, a constant, and given the normalization condition Eq.~\eqref{harmonicnorma} this constant is $1\over \sqrt{{\cal V}_{\cal N}}$, where ${\cal V}_{\cal N}$ is the volume of the manifold.  The co-exact zero-forms are the functions orthogonal to the constant function, and to label them we will generally use the index $a$ in place of the index $i_0$.  There are no exact zero-forms.  Our decomposition Eq.~\eqref{formdecompg} for functions then reads
\be \label{formdecompgscalar} \phi=\sum_{a} a^{a}\psi_{a}+{1\over \sqrt{{\cal V}_{\cal N}}}c^0~,
\ee
where $\psi^a$ is a complete orthonormal set of eigenvectors of the scalar Laplacian with positive eigenvalues,
\be \square\psi^a=-\lambda_a\psi^a~,\ \ \ \lambda_a>0~,\label{scalarprop1}\ee
\be \int d^Ny\sqrt{\gamma}\ \psi^{\ast}_{a}\psi_b=\delta^{}_{ab}~,\label{scalarprop2}\ee

For the case $p=1$, where we are considering covariant vectors $V_m$ on the manifold $\mathcal{N}$, we will often use $i$ in place of $i_1$ for the index labeling the co-exact vectors, and the label $\alpha$ instead of $\alpha_1$ for the harmonic vectors.  The decomposition Eq.~\eqref{formdecompg} then becomes
\begin{align}
V_m= \sum_i b^iY_{i,n}+\sum_{a}b^a \partial_n \psi_a+\sum_\alpha c^\alpha Y_{\alpha,n}~,
\end{align}
where $\psi_{a}$ is the orthonormal basis of positive eigenvalue eigenmodes of the scalar Laplacian (the same basis as appears in Eq.~\eqref{formdecompgscalar}), and
 $Y_{i,n}$ is a basis of orthonormal co-exact eigenvalues of the vector Laplacian, 
\bea && \Delta Y_{i,n} = -\square Y_{i,n}+R_n^{\ m}Y_{i,m} = \lambda_iY_{i,n}~,\ \ \ \lambda_i>0~, \\
&& \nabla^n Y_{i,n}=0~,\\
&& \int d^{N}y \sqrt{\gamma}\ Y_{i,m}^\ast   Y_{j}^m = \delta_{ij}~. 
\eea
The $Y_{\alpha,n}$ are the orthonormal harmonic forms, 
\bea && \Delta Y_{\alpha,n} = 0~, \\
&& \int d^{N}y \sqrt{\gamma}\ Y_{\alpha,n}^\ast  Y_{\beta}^n  = \delta_{\alpha\beta}~. 
\eea
As we will review in Appendix~\ref{appendixkilling}, in the case where the metric $\gamma$ is Einstein (the only case of interest in the Kaluza-Klein decompositions of gravity), harmonic vectors can only exist when the scalar curvature is non-positive, $R_{(N)}\leq 0$.  In this case the set indexed by $\alpha$ is empty when $R_{(N)}> 0$.

\subsection{Hodge Star\label{hodgeappendix}}

The Hodge star, $*$, maps $p$-forms to ($N-p$)-forms, 
\be (\ast \omega)_{n_{p+1}\cdots n _N}=\frac{1}{p!}\epsilon^{n _1\cdots n _p}_{\ \  \ \ \ \ \  n _{p+1}\cdots n _N}\omega_{n _1\cdots n _p}~,\ee
where $\epsilon_{n _1\cdots n _N}=\sqrt{\gamma}\tilde\epsilon_{n _1\cdots n _N}$ is the volume form and $\tilde\epsilon_{n _1\cdots n _N}$ is the anti-symmetric symbol with $\tilde\epsilon_{12\ldots N}=1$.  The volume form satisfies $ \epsilon^{p _1\cdots p _pn_1\cdots n_{d-p}}\epsilon_{p _1\cdots p _pm_1\cdots m_{d-p}}= p!(d-p)!\delta^{[n_1}_{m_1}\cdots \delta^{n_{d-p}]}_{m_{d-p}}.$

Inversely,
\be \omega_{n _1\cdots n _p}={1\over (N-p)!}\epsilon_{n _1\cdots n _p}^{\ \ \ \ \ \ \ n _{p+1}\cdots n _N}\left(\ast \omega\right)_{n _{p+1}\cdots n _N}~.\ee
We have $\ast\ast\omega=(-1)^{p(N-p)}\omega$, and the Hodge star is an isomorphism between the space of $p$-forms and the space of ($N-p$)-forms.

We can write the co-exterior derivative Eq.~\eqref{coextd} as $d^\dag=(-1)^{N(p+1)+1}\ast d\ast$, and so the Hodge star intertwines $d$ and $d^\dag$ and commutes with the Laplacian,
\be d\ast\omega=(-1)^{p}\ast d^\dag\omega~, \ \ \ d^\dag\ast\omega=(-1)^{p+1}\ast d\omega~, \ \ \ \ast \ \Delta=\Delta\ast~,\ee
for a $p$-form $\omega$.
In addition, it preserves the inner product,
\be (\ast\omega,\ast\eta)=(\omega,\eta)~,\ \ \ (\ast\alpha,\omega)=(-1)^{p(N-p)}(\alpha,\ast\omega)~,\ee
for $p$-forms $\omega$ and $\eta$ and an $(N-p)$-form $\alpha$.

 The Hodge star is an isomorphism between the space of harmonic $p$-forms and the space of harmonic ($N-p$)-forms
\be \Lambda^p_{\rm harm}(M) \leftrightarrow \Lambda^{N-p}_{\rm harm}(M)~.\ee
This means that the index $\alpha_p$ takes values over the same set as the index $\alpha_{N-p}$.  We can therefore choose our basis of harmonic ($N-p$)-forms to be the dual of the basis of $p$-forms,
\be Y_{\alpha_{N-p}}=\ast Y_{\alpha_p}~,\ \ \ 0\leq p< N/2~. \label{harmhodbasdu}\ee

In the case of even $N$, the space of harmonic $N/2$-forms is mapped to itself under the Hodge star.  For $N=4$ mod 4, we have $\ast^2=1$, so the space of harmonic $N/2$-forms splits into two eigenspaces of $\ast$ with eigenvalues $\pm 1$, the self-dual and anti-self dual forms, and we can choose our basis of $Y_{\alpha_{N/2}}$ to line up with this split.  We split the index set $\alpha_{N/2}$ into two sets, $\alpha_+$ and $\alpha_-$ such that $\ast Y_{\alpha_{\pm}}=\pm Y_{\alpha_{\pm}}$.  For $N=2$ mod 4, we have $\ast^2=-1$ and so the space of $N/2$-forms splits into two eigenspaces of $\ast$ with eigenvalues $\pm i$, the imaginary self-dual and imaginary anti-self dual forms, and we choose our basis to transform as $\ast Y_{\alpha_{\pm}}=\pm iY_{\alpha_{\pm}}$.  In this imaginary case we have $Y_{\alpha_+}^\ast=Y_{\alpha_-}$.  This case is useful for dealing with flux compactifications in $N=2$.

For any given eigenvalue $\lambda>0$, the Hodge star is an isomorphism between the space of exact $p$-forms of eigenvalue $\lambda$ and the space of co-exact ($N-p$)-forms of eigenvalue $\lambda$.
\be E^p_{\lambda,{\rm exact}}(M) \leftrightarrow E^{N-p}_{\lambda,{\rm co-exact}}(M)~, \ \ \   \lambda>0~.\ee
This means the index $i_p$ ranges over the same set as the index $i_{N-1-p}$.
We may choose our basis of forms so that
\be Y_{i_{N-1-p}} = \frac{1}{\sqrt{\lambda_{i_p}}}\ast dY_{i_p}~,\ \ \ 0\leq p\leq \lfloor N/2\rfloor~.\label{excohodrel}\ee
The normalization on the right hand side ensures that the left hand side is normalized properly under the scalar product Eq.~\eqref{formprod}.

These duality relations are important in the case of the flux compactifications of Section~\ref{fluxsection}.  For example, there can be harmonic one-forms in the expansion of the off-diagonal components of the higher dimensional graviton, corresponding to vectors in the uncompactified $d$-dimensional space.  In this case there are also harmonic ($N-1$)-forms coming from the expansion of the ($N-1$)-form potential with all indices in the compactified dimensions.  These harmonic ($N-1$)-forms correspond to scalars in the $d$ dimensions, but because of Hodge duality these scalars carry the same index, $\alpha$, as the harmonic one-forms from the graviton.  These scalars mix with the vectors coming from the graviton.
Similarly, there are co-exact vectors in the expansion of the off-diagonal components of the higher dimensional graviton, again corresponding to vectors in the uncompactified $d$-dimensional space with index $i$.  There are co-exact ($N-2$)-forms in the expansion of the ($N-1$)-form flux with one index in the $d$-dimensional space, corresponding to vectors in $d$ dimensions.  These co-exact ($N-2$)-forms are dual to the exact two-forms, and hence, because of the isomorphism Eq.~\eqref{bijectionddag}, carry the same index, $i$, as the co-exact one-forms.  These vectors mix with the vectors of the graviton.

\section{Killing Vectors on Closed Einstein Manifolds}\label{appendixkilling}

In the Kaluza-Klein decomposition of gravity, a special role is played by Killing vectors on the internal manifold, which end up corresponding to massless vector fields in the un-compactified space.  In this Appendix we review some needed facts about Killing vectors on closed Einstein manifolds.

We are considering a closed Riemannian $N$-manifold which is Einstein, i.e. the metric satisfies 
\be R_{mn}={R_{(N)}\over N}\gamma_{mn}~,\label{einsteincond}\ee
with $R_{(N)}$ constant.

A Killing vector $K^m$ is a vector along which the metric has an isometry,
\be \label{Killingequation} {\cal L}_K \gamma_{mn}\equiv \nabla_m K_n+\nabla_n K_m=0~.\ee

Taking the trace and divergence of Eq.~(\ref{Killingequation}), and using the Einstein condition Eq.~\eqref{einsteincond} we find that Killing vectors satisfy
\be \nabla_m K^m=0~, \ \ \ \ \square K^m+{R_{(N)}\over N} K^m=0~.\label{condone}\ee 

First, we prove some vanishing theorems which will tell us when Killing vectors can be present.  Consider the following identity for a general vector field $V_m$.
\be \nabla_m\left(V^n\nabla_n V^m-V^m\nabla_n V^n\right)=\nabla_m V_n\nabla^n V^m-\left(\nabla_m V^m\right)^2+{R_{(N)}\over N}V^2~.\ee
 Upon integrating both sides, the left hand side is zero because it is a total derivative, so from the right hand side we have
\be \int_{\cal N}\sqrt{\gamma}\left(\nabla_m V_n\nabla^n V^m-\left(\nabla_m V^m\right)^2+{R_{(N)}\over N}V^2\right)=0~.\label{genintvan}\ee

In the case where $V^m$ is a Killing vector $K^m$, we have, using the properties $\nabla_m K_n+\nabla_n K_m=0$ and $\nabla_m K^m=0$ from Eqs.~\eqref{Killingequation} and~\eqref{condone},
\be \int_{\cal N}\sqrt{\gamma}\left(-\nabla_m K_n\nabla^m K^n+{R_{(N)}\over N}K^2\right)=0~.\ee
Suppose $R_{(N)}$ is negative.  Then the integrand is a sum of squares and the integral cannot be zero unless $K^m=0$.  
\begin{itemize}
\item If $R_{(N)}<0$, there can be no non-trivial Killing vectors.  
\end{itemize}

This same trick allows us to prove a vanishing theorem for harmonic forms.  Consider Eq.~\eqref{genintvan} in the case where $V_m$ is a harmonic one-form $H_m$. We have, using the properties that a harmonic form is closed and co-closed, $\nabla_m H_n=\nabla_n H_m$ and $\nabla_m H^m=0$, 
\be \int_{\cal N}\sqrt{\gamma}\left(\nabla_m H_n\nabla^m H^n+{R_{(N)}\over N}H^2\right)=0~.\ee
Suppose $R_{(N)}$ is positive.  Then the integrand is a sum of squares and the integral cannot be zero unless $H_m=0$.  
\begin{itemize}
\item If $R_{(N)}>0$, there can be no non-trivial harmonic forms.  
\end{itemize}

In this sense, Killing vectors and harmonic forms are complementary.  They only co-exist in the case where $R_{(N)}=0$.  In this case, for the integral to be zero we see that both the Killing vectors and harmonic forms must satisfy $\nabla_m V_n=0$.  When $R_{(N)}=0$, the space of Killing vectors and the space of harmonic forms are identical.

The following integral is always greater than or equal to zero, and is equal to zero if and only if $V^m$ is a Killing vector,
\bea\label{Killingpositivity} 0&\leq &\int_{\cal N}\sqrt{\gamma}\left(\nabla_m V_n+\nabla_n V_m\right)\left(\nabla^m V^n+\nabla^n V^m\right)\\ &=&2\int_{\cal N}\sqrt{\gamma}\left[-V_m\square V^m-V^n\nabla_n\left(\nabla_m V^m\right)-{R_{(N)}\over N}V^2\right] \\ &=&2\int_{\cal N}\sqrt{\gamma}\ V_m\left( \Delta_K V\right)^m~,
\eea
where we have defined the following operator acting on the space of vector fields,
\be\left( \Delta_K V\right)^m=-\square V^m-\nabla^m\left(\nabla_n V^n\right)-{R_{(N)}\over N}V^{m}~.\ee
It is easy to check that this operator is elliptic, self-adjoint, and, because of Eq.~(\ref{Killingpositivity}), positive definite.  The Killing vectors are precisely the space of zero eigenvalues of this operator.  Combining with Eq.~\eqref{condone}, we have shown
\begin{itemize}
\item $K^m$ is a Killing vector if and only if \be\label{Killingconditions} \square K^m+{R_{(N)}\over N}K^{m}=0\ {\rm and}\ \nabla_m K^m=0~.\ee
\end{itemize}

Recalling the Hodge Laplacian acting on one-form fields, $\Delta V_ n=-\square V_ n+R_ n^{\ m}V_m,$
and using Eq.~(\ref{einsteincond}), we have 
\begin{itemize}
\item For $R_{(N)}>0$, the space of Killing vectors is precisely the space of co-exact eigenvectors of the vector Laplacian with eigenvalue $\lambda={2R_{(N)}\over N}$, \be\label{Killingconditionseinstein} \Delta K^ n=\left({2R_{(N)}\over N}\right)K^ n~,\ \ \ \nabla_ n K^ n=0~.\ee 
\end{itemize}
Recall that for $R_{(N)}=0$, the space of Killing vectors is precisely the space of harmonic vectors, and by the vanishing theorem above, there are no Killing vectors when $R_{(N)}<0$.  

From Eq.~(\ref{Killingpositivity}), positivity of the operator $-\square-{R_{(N)}\over N}$, we see that $\lambda={2R_{(N)}\over N}$ is in fact a lower bound on the possible eigenvalues of $\Delta$ on the space of co-exact forms.  It is saturated only for Killing vectors.

\section{Conformal Killing Vectors on Closed Riemannian Manifolds}\label{appendixconformal}

In the Kaluza-Klein decomposition of gravity, a special role is played by the so-called conformal scalars on the internal manifold.  These end up accounting for absent modes from the point of view of the un-compactified space, so it is important to understand when they are present.   In this Appendix we review some needed facts about conformal scalars on closed Einstein manifolds.

A conformal Killing vector $C^m$ is a vector along which the metric changes by an overall conformal factor,
\be \label{conformalKillingequation}{\cal L}_C \gamma_{mn}= \nabla_m C_n+\nabla_n C_m=f\, \gamma_{mn}~.\ee
where $f(y)$ is a scalar function.
Taking the trace of Eq.~(\ref{conformalKillingequation}), we find
\bea f={2\over N}\nabla_m C^m~,\eea 
so the conformal Killing equations Eq.~(\ref{conformalKillingequation}) can be written in the equivalent form
\be \label{conformalKillingequation2} \nabla_m C_n+\nabla_n C_m-{2\over N}\left(\nabla_p C^p\right) \gamma_{mn}=0~.\ee
The Killing vectors form a subspace of the conformal Killing vectors; they are precisely those for which $\nabla_m C^m=0$.  

In the following, we assume $N\geq 2$.  The case $N=1$ is trivial: every vector is a conformal Killing vector.

We start by proving some vanishing theorems.  The following integral is always greater than or equal to zero, and is equal to zero if and only if $V^m$ is a conformal Killing vector,
\bea\label{conformalKillingpositivity} \nn 0\leq &&\int_{\cal N}\sqrt{\gamma}\left(\nabla_m V_n+\nabla_n V_m-{2\over N}\left(\nabla_p V^p\right) \gamma_{mn}\right)\left(\nabla^m V^n+\nabla^n V^m-{2\over N}\left(\nabla_p V^p\right) \gamma^{mn}\right)\\\nn  =&&2\int_{\cal N}\sqrt{\gamma}\left[\nabla_n V_m\nabla^n V^m+{N-2\over N}\left(\nabla_m V^m\right)^2-{R_{(N)}\over N}V^2\right] \\ =&&2\int_{\cal N}\sqrt{\gamma} \ V_m \left( \Delta_C V\right)^m~,
\eea
where we have defined the following operator acting on the space of vector fields,
\be\label{conformalKillingoperator} \left( \Delta_C V\right)^m=-\square V^m-{N-2\over N}\nabla^m\left(\nabla_n V^n\right)-{R_{(N)}\over N}V^{m}~.\ee
It is easy to check that this operator is elliptic, self-adjoint, and, by Eq.~(\ref{conformalKillingpositivity}), positive.  The conformal Killing vectors are precisely the space of zero eigenvalues of this operator.

Suppose $R_{(N)}\leq 0$, and $V^m$ is a conformal Killing vector.   Looking at the middle line Eq.~(\ref{conformalKillingpositivity}), we see that the integrand is a sum of squares and the integral cannot be zero unless $V^m=0$.  If $R_{(N)}=0$, then we have $\nabla_m V_n=0$.  We thus have the following vanishing theorems:
\begin{itemize}
\item If $R_{(N)}<0$, there are no non-trivial conformal Killing vectors.  
\item If $R_{(N)}=0$, there are no conformal Killing vectors which are not also Killing vectors.
\end{itemize}

 Since there are no non-Killing conformal Killing vectors $R_{(N)}\leq 0$, we turn to the case $R_{(N)}>0$ for which they may exist.  
Taking the divergence of Eq.~(\ref{conformalKillingequation2}), we have
\be\label{cconformeq2} \square C^m+{R_{(N)}\over N}C^m+{N-2\over N}\nabla^m\left(\nabla_n C^n\right)=0~,\ee
which is the same as the operator Eq.~\eqref{conformalKillingoperator}, so we may write $\left( \Delta_C C\right)^m=0$.
Taking another divergence gives
\be \label{eigenvalue1} \Delta\left(\nabla_m C^m\right)={R_{(N)}\over N-1}\left(\nabla_m C^m\right)~, \ee
where $\Delta=-\square$ is the scalar Laplacian.  Thus $\nabla_m C^m$ is an eigenfunction of the Laplacian with eigenvalue $\lambda={R_{(N)}\over N-1}$.  In terms of this value of $\lambda$, we can rearrange derivatives in Eq.~(\ref{eigenvalue1}) to give
\be \nabla^m\left(\Delta C_m-\lambda C_m\right)=0~,\ee
where $\Delta$ is the Laplacian on vectors.
This implies 
\be \label{Cveceigenvalue} \Delta C_m=\lambda C_m+V^{\rm co-closed}_m~,\ee
where $V^{\rm co-closed}_m$ is an arbitrary co-closed form, $\nabla^m V^{\rm co-closed}_m=0$.  Now, decompose $C_m$ into its co-closed and exact parts,
\be \label{Cdecomposition} C_m=\nabla_m C+C_m^{\rm co-closed}~,\ee
where $C$ is a scalar with no constant component, and $\nabla^m C^{\rm co-closed}_m=0$.  Plugging Eq.~(\ref{Cdecomposition}) into Eq.~(\ref{Cveceigenvalue}), and reading off the co-closed and exact parts, we have
\be\label{Cdecomposition2} \nabla_m\left(\Delta C-\lambda C\right)=0~,\ \ \ \left(\Delta -\lambda\right) C_m^{\rm co-closed}=V^{\rm co-closed}_m~.\ee
The second equation in Eq.~(\ref{Cdecomposition2}) can be inverted for $C_m^{\rm co-closed}$ given $V_m^{\rm co-closed}$.  The first equation tells us that $C$ is an eigenvalue of the Laplacian up to a constant $C_0$, $\Delta C=\lambda C+C_0$, but since $C$ was assumed to have no constant piece, we must have $C_0=0$.  In summary, we have
\be \label{Cform} C_m=\nabla_m C+C_m^{\rm co-closed}~,\ \ \ \ \Delta C=\lambda C~,\ee
where $C_m^{\rm co-closed}$ is an arbitrary form satisfying $\nabla^m C_m^{\rm co-closed}=0$.  

Now, plugging Eq.~(\ref{Cform}) into Eq.~(\ref{cconformeq2}), we find $\square C_m^{\rm co-closed} +{R_{(N)}\over N}C_m^{\rm co-closed}=0$, which combined with the fact that $C_n^{\rm co-closed}$ is co-closed and the result Eq.~\eqref{Killingconditions}, tells us that $C_n^{\rm co-closed}$ is a Killing vector: $\nabla_m C_n^{\rm co-closed}+\nabla_n C_m^{\rm co-closed}=0$.   Therefore, the space of conformal Killing vectors is all vectors of the form
\be \nabla^m C+K^m~, \ \ \ \ \Delta C={R_{(N)}\over N-1} C~,\ee
where $K^m$ is an arbitrary Killing vector.  
\begin{itemize}
\item The subspace of conformal Killing vectors orthogonal to the subspace of Killing vectors is $\nabla^m C$, $\Delta C={R_{(N)}\over N-1} C$, i.e. the image under the gradient of the eigenspace of the scalar Laplacian with eigenvalue $\lambda={R_{(N)}\over N-1}$.  
\end{itemize}

These scalars, spanning the eigenspace of the scalar Laplacian with eigenvalue $\lambda={R_{(N)}\over N-1}$, whose gradients give the non-Killing conformal Killing vectors, are called \textit{conformal scalars}.  They must be treated specially in the decomposition of the graviton. 

\subsection{Lichnerowicz Bound\label{Lichboundap}}

A theorem of Lichnerowicz \cite{Lichnerowicz} says that given a closed $N$-dimensional Riemannian manifold $N\geq 2$, if the Ricci scalar $R_{(N)}$ satisfies $R_{(N)} \geq k$ for some constant $k > 0$, then the first nonzero eigenvalue $\lambda$ of the scalar Laplacian satisfies $\lambda \geq {k\over N-1}$.  Our interest is in Einstein spaces, for which $R_{(N)}$ is constant, so we have the bound
\be \lambda \geq {R_{(N)}\over N-1}~.\label{lichbound}\ee
Thus, the conformal Killing vectors of a closed Einstein manifold come precisely from the gradients of the lowest possible non-zero eigenfunctions of the scalar Laplacian, those that saturate the Lichnerowicz bound.

By a theorem of Obata \cite{Obata}, the equality in Eq.~\eqref{lichbound} is obtained if and only if the manifold is isometric to the sphere, so conformal scalars (and hence non-Killing conformal Killing vectors) exist only for the sphere.  The eigenvalues of the scalar Laplacian on an $N$ sphere of radius ${\cal R}$ are given by $l(l+N-1)/{\cal R}^2$, where $l=0,1,2,\ldots$, or in terms of the curvature $R_{(N)}=N(N-1)/{\cal R}^2$, 
\be \lambda_l={l(l+N-1)\over N(N-1)} R_{(N)}~.\ee
We see that the Lichnerowicz bound Eq.~\eqref{lichbound} is saturated by the eigenmodes with $l=1$.   In summary, conformal scalars exist only on the sphere, and are precisely the $l=1$ spherical harmonics.

Turning to the case $N=1$, we note that all scalars orthogonal to the constant scalar are conformal scalars.  In this case $R_{(N)}=0$, and the Lichnerowicz bound does not apply.

\section{Hodge Eigenvalue Decomposition for Symmetric Tensors}\label{appendixtensors}

In this Appendix we describe the analog of the Hodge decomposition for symmetric tensors.  This is needed to decompose the extra-dimensional components of the graviton.  The Killing vectors described in Appendix~\ref{appendixkilling} and the conformal scalars described in Appendix~\ref{appendixconformal} play a role in this expansion and must be treated with care.

Consider the space of (complex) symmetric tensors $h_{mn}$ on an Einstein space ${\cal N}$.  This space is denoted $S^2({\cal N})$.  We define the positive definite inner product
\be (h,h')=\int d^Ny\sqrt{\gamma}\ h_{mn}^\ast h'^{mn}~.\label{tensorinner}\ee

The natural Laplacian on this space is the Lichnerowicz operator \cite{Lichnerowicz1961}, defined by 
\be \Delta_L h_{mn}=-\nabla^2 h_{mn}+{2R_{(N)}\over N}h_{m n}-2R_{mpnq}h^{pq}~.\label{lichoperatorA}\ee
The Lichnerowicz operator is self-adjoint with respect to the inner product Eq.~\eqref{tensorinner}, $(h, \Delta_Lh')=( \Delta_Lh,h'),$ thus we can decompose the space of symmetric tensors into eigenspaces of the Lichnerowicz operator,
\be \label{Lichnerowiczeigenspaces} S^2({\cal N})=\sum_{\oplus\lambda}S^2_\lambda({\cal N})~.\ee
where $E_\lambda^p({\cal N})$ are the subspaces $\{\omega\in \Lambda^p({\cal N})|\Delta\omega=\lambda\omega\}$.  Each subspace is finite dimensional.  We only consider those $\lambda$'s such that the subspaces are non-trivial, and this forms the spectrum of the Lichnerowicz Laplacian.  Due to self-adjointness, the $\lambda$'s are all real and the eigenspaces for different $\lambda$ are orthogonal with respect to the inner product Eq.~\eqref{tensorinner}.  

The space of symmetric tensors can also be decomposed as the orthogonal sum of the space of traverse tensors $S^2_T({\cal N})$, i.e. those tensors $h^T_{mn}$ satisfying $\nabla^m h^T_{mn}=0$, and all tensors of the form $\nabla_{(m}V_{n)}$ for some vector $V_m$  \cite{besse2007einstein,Ishibashi:2004wx}
\be S^2({\cal N})=S^2_T({\cal N})\oplus {\rm Im}\(\nabla_{(m}\bullet_{n)}\)~.\label{Tdiffdec}\ee  
The vector can in turn be split into transverse and longitudinal parts according to the standard Hodge decomposition Eq.~\eqref{hodgedecomposition}: $V_m=\partial_m\phi+V^T_m$, $\nabla^m V^T_m=0$, where $\phi$ is some scalar.  Finally we extract the trace, with the result that we can write any symmetric tensor as 
\be h_{mn}=h_{mn}^{TT}+2\nabla_{(m}V^T_{n)}+2\(\nabla_m\nabla_n\phi-{1\over N}\nabla^2\phi\) + {1\over N}\bar\phi \, \gamma_{mn}~, \label{tensorhodge}\ee
where $h_{mn}^{TT}$ is transverse and traceless: $\nabla^m h_{mn}^{TT}=0$, $\gamma^{mn}h_{mn}^{TT}=0$, and $\bar\phi$ is another scalar, which carries the trace: $\bar \phi=h$. This is the Hodge decomposition for symmetric tensors.  The four parts of this decomposition are all orthogonal with respect to the inner product Eq.~\eqref{tensorinner}.

The Lichnerowicz operator Eq.~\eqref{lichoperatorA} commutes with traces, divergences and symmetrized derivatives,
\bea 
\gamma^{mn}\Delta_L h_{mn}&=&\Delta(\gamma^{mn} h_{mn})~,\\
\Delta_L\( \phi\, \gamma_{mn}\)&=&\gamma_{mn}\Delta\phi~,\\
\Delta_L \(\nabla_{(m}V_{n)}\)&=&\nabla_{(m}\Delta V_{n)}~, \\
\nabla^m \Delta_L h_{mn}&=&\Delta\(\nabla^m h_{mn}\)~, 
\eea
where $V_m$ and $\phi$ are any one-form and scalar, and $\Delta$ is the Hodge Laplacian Eq.~\eqref{Hodgelapgen} (we also have $\nabla^m \Delta V_m=\Delta\(\nabla^m V_m\)$).

In light of these relations, the decomposition Eq.~\eqref{Lichnerowiczeigenspaces} according to Lichnerowicz eigenvalues commutes with the tensor Hodge decomposition Eq.~\eqref{tensorhodge}.  This means we can write an arbitrary symmetric tensor $h_{mn}$ as
\bea \label{tensordecompc} h_{mn}&=&\sum_{\cal I} c^{\cal I}h_{mn,{\cal I}}^{TT}+\sum_{i\not= {\rm Killing}}c^i\left(\nabla_m \xi_{n,i}+\nabla_n \xi_{m,i}\right)+\sum_{a\not={\rm conformal}} c^a\left(\nabla_m\nabla_n\psi_a-{1\over N}\nabla^2\psi_a \gamma_{mn}\right)\nn\\ && +\sum_a {1\over N} \bar c^a\psi_a \gamma_{mn}+{1\over N}c^0 \gamma_{mn}~.
\eea
The various $c$'s are all coefficients, the generalized Fourier coefficients; they are uniquely determined by and determine $h_{mn}$.  The different parts of this expression are as follows: $h_{mn,{\cal I}}^{TT}$ are a complete orthonormal basis, indexed by ${\cal I}$,  of transverse traceless tensors, chosen so that they are eigenvalues of the Lichnerowicz operator with eigenvalues $\lambda_{\cal I}$,
\bea &&\Delta_L h_{mn,{\cal I}}^{TT}=\lambda_{\cal I} h_{mn,{\cal I}}^{TT}~,\ \ \ \ \ \nabla^mh_{mn,{\cal I}}^{TT}=\gamma^{mn}h_{mn,{\cal I}}^{TT}=0~,\\
&&\int d^Ny\sqrt{\gamma}\, \(h^{TTmn,{\cal I}}\)^\ast h_{mn,{\cal J}}^{TT}=\delta^{{\cal I}}_{\cal J}~.
\eea
The $\xi_{m,i}$ are a complete orthonormal basis, indexed by $i$,  of co-closed one-forms, chosen so that they are eigenvalues of the Hodge Laplacian with eigenvalues $\lambda_i$,
\bea &&\Delta \xi_{m,i}=\lambda_i \xi_{m,i}~,\ \ \ \ \nabla^m\xi_{m,i}=0~,\\
&&\int d^Ny\sqrt{\gamma}\, \(\xi^{m,i}\)^\ast \xi_{m,j}=\delta^{i}_j~.
\eea
In the sum over $i$ in Eq.~\eqref{tensordecompc}, we omit those with eigenvalue $\lambda={2R_{(N)}\over N}$, because these are precisely those that are Killing vectors (see Appendix~\ref{appendixkilling}), i.e. $\nabla_m\xi_n+\nabla_n\xi_m=0$ (and hence these do not get their own independent Fourier coefficient $\phi^i$).  The $\psi^a$ are a complete orthonormal basis of the scalar Laplacian, as in Eqs.~\eqref{formdecompgscalar},~\eqref{scalarprop1},  and~\eqref{scalarprop2}.  In the sum $\sum_{a\not={\rm conf.}}$, we have indicated that we should leave out the conformal scalars (see Appendix~\ref{appendixconformal}), those scalars with eigenvalue $\lambda={R_{(N)}\over N-1}$, because these are precisely those whose gradients are conformal Killing vectors, for which $\nabla_m\nabla_n\psi-{1\over N}\nabla^2\psi \gamma_{mn}=0$, and hence should not get their own Fourier coefficient $c^a$.  
 
 \subsection{\label{ModspaceA}Moduli Space of Einstein Structures}
 
 In the decomposition of the graviton, massless scalars appear corresponding to directions in the moduli space of Einstein structures of the internal manifold.
The moduli space of Einstein structures on a manifold is the space of all possible Einstein metrics which can be put on the manifold, modulo diffeomorphisms, and with the condition that the total volume be fixed \cite{besse2007einstein}.  

Considering small variations of the metric $\delta \gamma_{mn}=h_{mn}$, we see from Eq.~\eqref{Tdiffdec} that the condition that the variation not be a diffeomorphism is the condition that $h_{mn}$ be transverse, $\nabla^m h_{mn}=0$.  The condition that the volume be fixed is the condition $\delta \int d^Ny \sqrt{\gamma}={1\over 2} \int d^Ny \sqrt{\gamma}\gamma^{mn}h_{mn}= 0$, i.e. that the integral of the trace, $h\equiv \gamma^{mn}h_{mn}=0$, vanish.  

Plugging the general decomposition \eqref{tensordecompc} into the equation $\nabla^m h_{mn}=0$, we find that the most general transverse tensor can be written as
\bea  h_{mn}&=&\sum_{\cal I} c^{\cal I}h_{mn,{\cal I}}^{TT}+\sum_{a\not={\rm conformal}} c^a\left(\nabla_m\nabla_n\psi_a+\left(\lambda_a-{R_{(N)}\over N}\right)\psi_a \gamma_{mn}\right)+{1\over N}c^0 \gamma_{mn}~.\nn\\ \label{tensordecompctransve}
\eea

The condition that a space be Einstein can be written $R_{mn}-{R_{(N)}\over N}\gamma_{mn}=0.$  Taking the variation of this Einstein condition, and plugging in \eqref{tensordecompctransve} for the variation, we have
\bea 2\delta\( R_{mn}-{R_{(N)}\over N}\gamma_{mn}\)=&& \sum_{\cal I} c^{\cal I}\left(-\nabla^2 h_{mn,{\cal I}}^{TT}-2R_{mpnq}h^{ pq\, TT}_{\cal I}\right) \nn\\ &&-\sum_{a\not={\rm conformal}} c^a (N-2)  \left( \lambda_a-{R_{(N)}\over N}\right) \left(\nabla_m\nabla_n\psi_a-{1\over N}\nabla^2\psi_a \gamma_{mn}\right) .~\nn\\  \label{einsteinfvneedtovanishe}\eea
Looking at Eq.~\eqref{lichoperatorA}, we can rewrite the first term,
 \be -\nabla^2 h_{mn,{\cal I}}^{TT}-2R_{mpnq}h^{ pq\, TT}_{\cal I}=\left(\Delta_L - {2R_{(N)}\over N}\right)h_{mn,{\cal I}}^{TT}=\left(\lambda_{\cal I} - {2R_{(N)}\over N}\right)h_{mn,{\cal I}}^{TT}.\ee
For \eqref{einsteinfvneedtovanishe} to vanish, each coefficient must individually vanish since it is in the form of the general expansion \eqref{tensordecompc}.  Thus we must have $c^a=0$ when $N>2$ (the coefficient $\left( \lambda_a-{R_{(N)}\over N}\right) $ is never zero because of the Lichnerowicz bound Eq.~\eqref{lichbound}), and we must have $c^{\cal I}=0$, except for the case $\lambda_{\cal I}={2R_{(N)}\over N}$ when the coefficient vanishes.  Finally, the condition that the volume be fixed, $\int d^Ny \sqrt{\gamma}\gamma^{mn}h_{mn}= 0$, fixes $c^0=0$.  

Thus, we see that when $N>2$ the tangent space to the moduli space, at a given Einstein metric, is spanned precisely by transverse traceless tensors which are eigentensors of the Lichnerowicz operator with eigenvalue ${2R_{(N)}\over N}$,
\be \Delta_L h^{TT}_{mn}={2R_{(N)}\over N}h^{TT}_{mn}~.\ee
These give rise to massless scalar states in the Kaluza-Klein reduction of the graviton.

\bibliographystyle{utphys}
\addcontentsline{toc}{section}{References}
\bibliography{KKv2}

\end{document}